\documentclass[11pt]{article}

\pdfoutput=1

\usepackage[utf8]{inputenc}

\usepackage[top=1in,bottom=1in,left=1in,right=1in]{geometry}
\usepackage{graphicx}
\usepackage{epsfig}
\usepackage{amsfonts}
\usepackage{latexsym}
\usepackage{amsmath,amssymb}
\usepackage{subfig}
\usepackage{xcolor}
\usepackage{physics}
\usepackage{tikz}
\usepackage{stmaryrd}
\usepackage{hyperref}

\def\p{\partial}
\newcommand{\be}{\begin{equation}} 
\newcommand{\ee}{\end{equation}}

\usepackage{cleveref}
\crefname{equation}{Eq.}{Eqs.} 
\crefrangelabelformat{equation}{(#3#1#4--#5#2#6)}
\crefname{section}{§}{§§}
\Crefname{section}{§}{§§}

\numberwithin{equation}{section}

\newcommand{\q}{n_e^\vee}
\newcommand{\s}{n_m^\vee}

\begin{document}
\begin{titlepage}
\unitlength = 1mm
~ \vskip 3cm
\begin{center}

{\LARGE{{\textsc{Chiral Soft Algebras for $\mathcal{N}=2$ Gauge Theory}}}}

\vspace{1.25cm}
Erin Crawley, Andrew Strominger, and Adam Tropper

\vspace{.5cm}

{\it  Center for the Fundamental Laws of Nature,\\ Harvard University,
Cambridge, MA 02138}\\ 

\vspace{0.8cm}

\begin{abstract} 
Some time  ago, Seiberg and Witten solved for moduli spaces of vacua parameterized by scalar vacuum expectation values  in $\mathcal{N}=2$ gauge theories. More recently, new vacua associated to soft theorems and asymptotic symmetries have been found. This paper takes some first steps towards a complete picture of the infrared geometry of $\mathcal{N}=2$ gauge theory incorporating both of these infrared structures.

\end{abstract}

\vspace{1.0cm}

\end{center}

\end{titlepage}

\tableofcontents

\newpage

\section{Introduction}
In the space of all quantum field theories, four-dimensional $\mathcal{N}=2$ gauge theories occupy a special sweet spot. They are complicated enough to exhibit a wealth of subtle physical phenomena including confinement, monopole condensation and $S$-duality and  yet, as shown by Seiberg and Witten \cite{seiberg1994electric},  simple and tractable enough that many aspects are exactly soluble in the deep infrared. 
The precise mathematical formulae obtained in  \cite{seiberg1994electric}  along with later works  reviewed in \cite{tachikawa2013n, labastida2005topological, Bilal:1995hc} have had profound implications across disparate areas of mathematics and physics.
 
In the last decade, qualitatively new  phenomena have been discovered  in the deep infrared of gauge theories $e.g.$ \cite{Strominger:2013lka, He:2014cra, He:2015zea,Strominger:2017zoo,guevara2021holographic,strominger2021w}. These include an infinite tower of  soft theorems which are equivalent to spontaneously broken asymptotic symmetries.  The most compact representation of this phenomenon is in terms of {\it chiral soft algebras} generated by higher spin currents  acting holomorphically on the celestial sphere at null infinity. These imply an infinite-dimensional vacuum moduli space.\footnote{A simple  picture has emerged for the large gauge vacua associated to the leading soft theorem and spin one current \cite{Strominger:2017zoo},  while aspects of those associated to the subleading soft theorems and higher spin currents are less-understood.}Roughly speaking, the vacua can be thought of as differing by large gauge transformations or, equivalently, the addition of soft photons.

A complete understanding  of the  infrared structure of $\mathcal{N}=2$ gauge theories would 
incorporate both the well-understood older insights and the less-understood newer ones.  Integrating chiral soft algebras into  the concise and constrained mathematical structure of the $\mathcal{N}=2$ theories may also serve to sharpen our understanding of them. 
Many questions about how the old and new insights  combine arise. An immediate one is:
\begin{center}
{\it What is the complete moduli space of $\mathcal{N}=2$ gauge theory vacua?}
\end{center}
\noindent While subtle and challenging, this question might eventually be  answerable. We will not do so in this paper.  Rather we simply lay some  basic 
groundwork and make a few relevant new observations which point to a rich combined infrared structure.  

The complete moduli space, denoted   $\mathcal{M}$,  is infinite-dimensional, but of course must contain the familiar finite-dimensional space $\mathcal{M}_{\phi} \subset \mathcal{M}$ of massless scalar vevs.  
In this paper we  show, focusing on  $\mathcal{N}=2$ effective theories with a single $U(1)$ vector multiplet in the infrared, that the commutator of two  gauge generators (constructed from subleading soft photons) of the chiral soft algebra  gives motion on the scalar moduli space $\mathcal{M}_{\phi}$. This indicates that $\mathcal{M}_{\phi}$ is non-trivially embedded in $\mathcal{M}$. Were there no  irrelevant terms in the effective action, the chiral soft  algebra would be  abelian and the embedding trivial. However the algebra  is deformed by irrelevant dimension five $\phi F^2$ couplings whose presence is implied by the curvature of $\mathcal{M}_{\phi}.$  We compute explicitly the resulting supersymmetric non-abelian chiral soft algebra, which bears a strong resemblance to the (super loop group of the wedge of) $w_{1+\infty}$ appearing in gravity \cite{strominger2021w}. The structure constants vary over $\mathcal{M}_{\phi}$ and degenerate at singular points.\footnote{Related and relevant discussions of chiral soft algebras and moduli spaces appear in \cite{Kapec:2022hih, Kapec:2022axw}.}

In a parallel vein, the  structure of   $\mathcal{M}$ is in  part encoded in the variation of the soft S-matrix over  $\mathcal{M}_{\phi}$.
We  show that the electromagnetic tree-level leading soft S-matrix  is encoded in  2D celestial CFT (CCFT) correlators  of free  bosons whose target is a torus defined by  a $(2,2)$ Narain lattice. 
The moduli of this lattice is a section of an $SL_2(\mathbb{Z})$ bundle over $\mathcal{M}_{\phi}$. We find the lattice degenerates in an interesting fashion at singular points of $\mathcal{M}_{\phi}$, suggesting an enhanced chiral soft algebra.

This paper is organized as follows. In Section \ref{sec:review}, we give brief reviews of relevant material from $\mathcal{N} = 2$ effective field theory and chiral soft  algebras. 

In Section \ref{sec:holomorphicOPEs_woSings}, we begin by constructing the action of 4D $\mathcal{N} = 2$  supersymmetry on the 2D $SL_2(\mathbb{C})$ conformal primaries. It is an exotic realization of  2D  supersymmetry which  involves shifts in the conformal weight and no 2D derivatives. We then derive explicit formulae for  the chiral soft algebra  which turn out to be nonabelian even for  a single $U(1)$ vector multiplet. It is generated by rescaled conformal primaries with quantized weights $\Delta=1, \frac{1}{2},0, -\frac{1}{2},...$. The entire chiral soft algebra is shown to be $SL_2(\mathbb{Z})$ covariant, with bulk supercharges acting  as outer derivations. We then consider the addition  of a massless hypermultiplet, which may eventually be relevant for analysis of singularities in $\mathcal{M}_{\phi}$, and find that they give a further enlargement of the chiral soft algebra. 

In Section \ref{sec:Goldstones} we turn to  the electromagnetic part of the soft S-matrix as a function on $\mathcal{M}_{\phi}.$  In a conformal basis this is generated by a pair of 2D scalars corresponding to the Goldstone bosons of large gauge symmetry and the bosonized soft current, or soft boson. Exponentials of the Goldstone boson are Wilson-`t Hooft  lines which pierce the celestial sphere. Exponentials of soft bosons generate large gauge transformations. We show that mutual locality on the celestial sphere together with electromagnetic charge quantization force the 2D bosons to live on a torus  defined by a $(2,2)$ Narain lattice whose modular parameter is $\tau=\frac{\theta}{2\pi}+\frac{4\pi i}{e^2}$. The conformal primary exponentials have  integer spins and imaginary conformal weights.  Analytically  continuing from Minkowski to signature $(2,2)$ Klein space, we find that the conformal weights are real. Continuation to Klein space is natural because it is  implicit in the holomorphic limits used to define the chiral soft algebras. A relevant and  interesting  study of $\mathcal{N}=2$ gauge theory  in Klein space and its embedding in string theory can be found in \cite{Dijkgraaf:2016lym}. It was shown that central constructs  such  as Seiberg-Witten curves, Nekrasov's instanton calculus, D-branes  and geometric engineering  may all be harmoniously analytically continued to split signature.

Section \ref{sec:strong coupling} comments on the singular behavior of both the chiral soft algebra and the soft S-matrix at singular points of $\mathcal{M}_{\phi}$.

\section{Review and Preliminaries} \label{sec:review}
\subsection {Review of $\mathcal{N} = 2$  Gauge Theory}
\label{sec: SW theory review}

This subsection  reviews aspects of  the low-energy effective action for gauge theories with $\mathcal{N} = 2$ supersymmetry and describes their moduli space of  vacua $\mathcal{M}_{\phi}$ parameterized by scalars in vector multiplets. We suppress here the effects of large gauge transformations which generate the much larger vacuum  moduli space $\mathcal{M}$ of which 
$\mathcal{M}_{\phi}$ is a submanifold. A general $\mathcal{N} = 2$ supersymmetric theory is described by $v$  vector multiplets and $h$ hypermultiplets. Here we largely restrict to the simplest, but already very rich, case  $v = 1$ and $h = 0$ which arises as the low energy effective action of super Yang-Mills with gauge group $SU(2)$ considered  in \cite{seiberg1994electric}.

\subsubsection {$\mathcal{N}=2$ Effective  Actions }

The dynamical field content in a single-vector  effective theory includes a complex scalar, $\phi$, a $U(1)$ gauge field, $A_{\mu}$, and two Weyl spinors, $\psi_{I}^{\alpha}$ ($I = 1,2$). Note that while the UV theory has an $SU(2)$ symmetry, the gauge symmetry in the effective theory is spontaneously broken to $U(1)$ when the scalar field acquires a vacuum expectation value (vev). To this end, we write
\begin{equation}
    \phi=a+\varphi \quad \text { where } \quad a=\langle\phi\rangle\hspace{2pt},
    \label{eq: scalar field vev expansion}
\end{equation}
where $\varphi$ encodes fluctuations around the scalar's vev, $a$. In addition, there are non-propagating auxiliary fields, $F$ and $D$ appearing in the Lagrangian which may be integrated out.

 It will be convenient to package $\phi, \psi_{1}^{\alpha}$ and $F$ into an $\mathcal{N}=1$ chiral superfield, $\Phi$, and to package $A_{\mu}, \psi_{2}^{\alpha}$, and $D$ into another $\mathcal{N}=1$ superfield $W^{\alpha}$ defined by (in standard notation):
\begin{equation}
\begin{aligned}
\Phi(x, \theta, \bar{\theta}) & =\phi(y)-\sqrt{2} \theta \psi_{1}(y)+\theta \theta F(y)\hspace{2pt}, \\
W^{\alpha}(x, \theta, \bar{\theta}) & =-i \psi_{2}^{\alpha}(y)+\theta^{\alpha} D(y)-i\left(\sigma^{\mu \nu} \theta\right)^{\alpha} F_{\mu \nu}(y)+\theta \theta\left(\sigma^{\mu} \partial_{\mu} \bar{\psi}_{2}\right)^{\alpha}(y)\hspace{2pt},
\end{aligned}
\end{equation}
where $F_{\mu\nu} = \partial_\mu A_\nu - \partial_\nu A_\mu$ and we have introduced the two-component Grassmann spinors $\left\{\theta^{\alpha}, \bar{\theta}_{\dot{\alpha}}\right\}$ and the variable $y^{\mu}=x^{\mu}+i \theta \sigma^{\mu} \bar{\theta}$ which appear in the formulation of $\mathcal{N}=1$ superspace. The most general low-energy effective Lagrangian (i.e. one which contains no more than two derivatives) that retains the $\mathcal{N}=2$ supersymmetry is
\begin{equation}
\mathcal{L}_{\text {eff }}=\frac{1}{4 \pi} \operatorname{Im}\left[\int d^{2} \theta d^{2} \bar{\theta} \hspace{2pt}\frac{\partial \mathcal{F}(\Phi)}{\partial \Phi} \overline{\Phi}+\frac{1}{2} \int d^{2} \theta \hspace{2pt} \frac{\partial^2\mathcal{F}(\Phi)}{\partial \Phi^2} W^{\alpha} W_{\alpha}\right]\hspace{2pt},
\label{eqn: effective action N=1}
\end{equation}
where $\mathcal{F}(\Phi)$ is a holomorphic function (depending only on $\Phi$ and not $\overline{\Phi}$) called the symplectic prepotential. This Lagrangian is extremely constrained by the $\mathcal{N}=2$ supersymmetry -- indeed, it is completely fixed up to a  choice for the holomorphic symplectic prepotential  $\mathcal{F}(\Phi)$.

It is convenient to define the complex coupling coupling constant, $\tau(\phi)$, in terms of this prepotential by
\begin{equation}
\tau(\phi)\equiv \frac{\partial^2 \mathcal{F}(\phi)}{\partial \phi^2}.
\end{equation}  
The complex coupling $\tau$ is related to the gauge coupling constant and theta parameter by \\ $\tau =\frac{\theta}{2\pi} + i\frac{4\pi}{e^2}$. The Lagrangian \eqref{eqn: effective action N=1} has the following expression in terms of component fields \cite{labastida2005topological}\footnote{In this expression, $\tilde{F}^{\mu \nu}=\frac{1}{2} \varepsilon^{\mu \nu \rho \sigma} F_{\rho \sigma}$ is the dual field strength of $F_{\mu \nu}$ and $D^{IJ}$ is a matrix composed of the Lagrange multipliers
\begin{equation}
D^{IJ}=\begin{pmatrix}
\sqrt{2} F & i D\\
i D & \sqrt{2} F^{\dagger}\\
\end{pmatrix}.
\end{equation}}
\begin{equation}
\begin{aligned}\label{eq:Leff_component_fields}
\mathcal{L}_{\text {eff }}=-\frac{1}{4 \pi} & \operatorname{Im}\left[\tau(\phi)\left(\frac{1}{4} F^{\mu \nu} F_{\mu \nu}-\frac{i}{4} F^{\mu \nu} \widetilde{F}_{\mu \nu}+\partial_{\mu} \phi \partial^{\mu} \bar{\phi} + i \psi^{I} \sigma^{\mu} \partial_{\mu} \bar{\psi}_{I}\right)\right] \\
& +\frac{\sqrt{2}}{16 \pi} \operatorname{Im}\left[\tau^{\prime}(\phi)\left(\varepsilon^{IJ} \psi_{I} \sigma^{\mu \nu} \psi_{J} F_{\mu \nu}-\psi_{I} \psi_{J} D^{I J}\right)\right] \\
& +\frac{1}{16 \pi} \operatorname{Im}\left[\tau(\phi) D_{KL} D^{KL}\right]+\frac{1}{48 \pi} \operatorname{Im}\left[\tau^{\prime \prime}(\phi) \varepsilon^{IK} \varepsilon^{JL}\left(\psi_{I} \psi_{J}\right)\left(\psi_{K} \psi_{L}\right)\right]\hspace{2pt},
\end{aligned}
\end{equation}
where $\phi^\dagger = \bar{\phi}$ and $\psi_I^\dagger = \bar{\psi}^I = \varepsilon^{IJ} \bar{\psi}_J$ where $\varepsilon^{IJ}$ is the totally antisymmetric tensor (with $\varepsilon^{12} = \varepsilon_{21} = 1$ and $\varepsilon_{IJ} \varepsilon^{JK} = \delta^K_I$) responsible for raising and lowering the $\mathfrak{su}(2)$ R-symmetry indices $I,J$.

\subsubsection {The Moduli Space $\mathcal{M}_{\phi}$}
\label{sec: moduli space review}

The scalar field $\phi$ plays a privileged role in this analysis because different values of the scalar vev, $a=\langle\phi\rangle$ will generally encode inequivalent vacua, each preserving the $\mathcal{N}=2$ supersymmetry. The space of such vacua  is called the Coulomb branch of the moduli space and  denoted $\mathcal{M}_{\phi}$. From Equation \eqref{eq:Leff_component_fields}, one can read off the metric on $\mathcal{M}_{\phi}$ from the kinetic terms
\begin{equation}
d s^{2}=\frac{\operatorname{Im} \tau(a)}{4 \pi} \hspace{2pt} d a \hspace{1pt} d \bar{a}\hspace{2pt}.
\label{eqn: metric}
\end{equation}
Defining the so-called dual coordinate, $a_{D}$, via
\begin{equation}
a_{D}\equiv \frac{\partial \mathcal{F}(a)}{\partial a}\hspace{2pt},
\end{equation}
we see that $\mathcal{M}_{\phi}$ is a Kähler manifold with Kähler potential $K = \frac{1}{8\pi i}(\bar{a} a_{D}-a \bar{a}_D)$.

\begin{figure}
    \centering
    \resizebox{0.75\textwidth}{!}    {
        \input{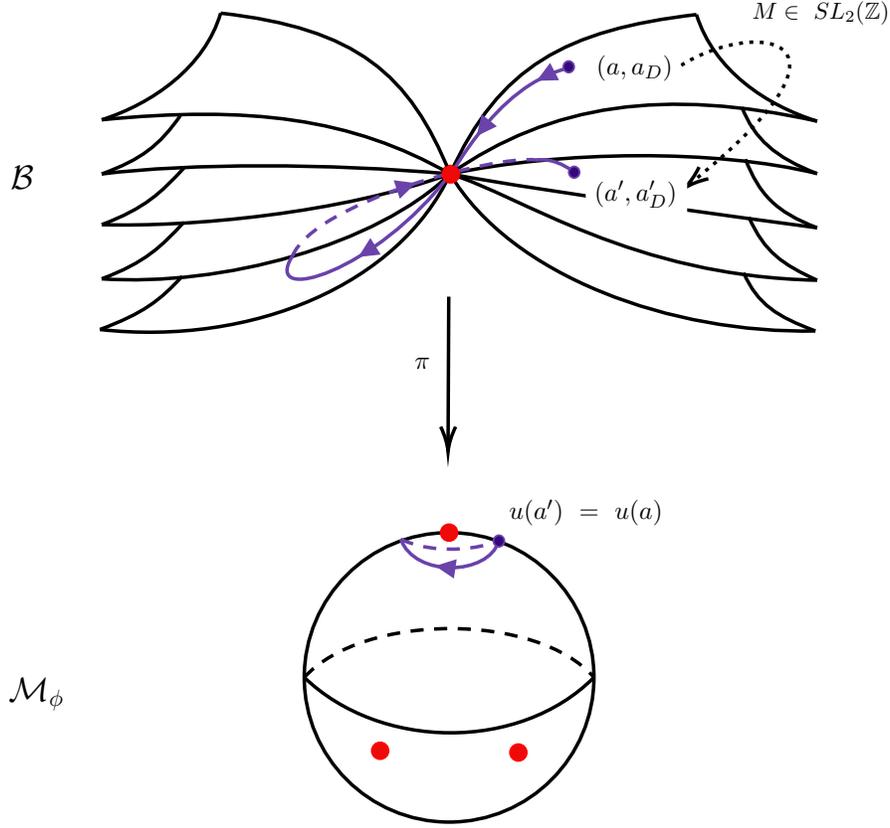}
    }
    \caption{The vacuum moduli space $\mathcal{M}_{\phi}$ and the $SL_2(\mathbb{Z})$ bundle $\mathcal{B}.$ A closed loop around a singularity in $\mathcal{M}_{\phi}$ induces a monodromy in $\mathcal{B}$ where the endpoint $(a',a_D') \in \mathcal{B}$ is related to $(a,a_D) \in \mathcal{B}$ by an element of $SL_2(\mathbb{Z}).$ } 
    \label{fig: bundle}
\end{figure}

Distinct choices for $a$ are not necesssarily inequivalent vacua. In particular, there is a $\mathbb{Z}_{2}=\text{Weyl}(\mathfrak{su}(2))$ gauge symmetry mapping $a \mapsto-a$, rendering these vacua secretly equivalent. One accordingly  defines the Weyl-invariant parameter
\begin{equation}
    u=\frac{1}{2}\left\langle\phi^{2}\right\rangle\hspace{2pt},
\end{equation}
which provides a good coordinate on  $\mathcal{M}_{\phi}$ in a neighborhood of  $u=0$. In the particular case of $SU(2)$ Seiberg-Witten theory, the $u$ coordinate ranges over the Riemann sphere with three punctures, which one may choose to be located at $u=1,-1$, and $\infty$. These punctures coincide with various singularities in the moduli space, and the aforementioned metric becomes singular at precisely these points. The singularity at $u=\infty$ is a weak coupling singularity where the theory becomes free. The singularities at $u=\pm$ 1 are strong coupling singularities due to massive hypermultiplets (a monopole at $u = 1$ and a dyon at $u = -1$) becoming massless and ruining the effective description.

\subsubsection {Action of $SL_{2}(\mathbb{Z})$ } \label{sec: SL(2, Z) review}

Low energy $\mathcal{N}=2$ effective theories have discrete families of redundant descriptions related by $SL_2(\mathbb{Z})$ transformations of the field variables.
The  $SL_2(\mathbb{Z})$ maps the pair of coordinates $\left(a_{D}, a\right)$ to the pair $\left(a_{D}', a'\right)$ related by
\begin{equation}
\binom{a_D}{a} \longmapsto \binom{a_D'}{a'}=\binom{w \hspace{10pt} x}{y \hspace{10pt} z}\binom{a_D}{a}\quad \text { with } \quad \binom{w \hspace{10pt} x}{y \hspace{10pt} z}  \in S L_{2}(\mathbb{Z})\hspace{2pt}.
\label{eqn: SL(2,Z) transform}
\end{equation}
Thus, if $\left(a_{D}, a\right)$ describe some point $u \in \mathcal{M}_{\phi}$, then so too do all other pairs $\left(a_{D}', a'\right)$. In this way, one can think of $(a_D,a)$ as living on an $SL_2(\mathbb{Z})$ bundle $\mathcal{B} \rightarrow \mathcal{M}_{\phi}$. As such, the various Lagrangian descriptions of the theory -- which are parameterized by $a$ -- should also be thought of as living on $\mathcal{B}$ (see Figure \ref{fig: bundle}). The duality transformation acts on $\tau$ via M\"obius transformation
\begin{equation}
    \tau \longmapsto \frac{w \tau + x}{y \tau + z}\hspace{2pt}.
\end{equation}

\subsection {Chiral soft algebras}
\label{sec: CCFT review}
Here we recap a few basic formulae describing conformal primary operators and the chiral soft algebras they generate \cite{Strominger:2017zoo, Raclariu:2021zjz, Pasterski:2021rjz, Pasterski:2021raf, McLoughlin:2022ljp}. 

\subsubsection{Conformal Primary Operators}
Let $X$ denote a particular species of massless particle with helicity $s_X$. We parameterize the momentum of this particle by an overall scale $\omega \in \mathbb{R}_+$ and complex coordinates $(z,\bar{z})$ on the celestial sphere at null infinity as
\begin{equation}
    p^\mu = \frac{\omega}{2}\big(1+z \bar{z},z+\bar{z},-i(z-\bar{z}),1-z\bar{z}\big)\hspace{2pt}.
\end{equation}
From the 4D annihilation operator $a_{X}(\omega,z,\bar z)$ for this particle (see Appendix A), one defines the corresponding 2D outgoing operator $\mathcal{O}^{X}_\Delta(z,\bar z)$ in the  CCFT via the Mellin transform 
\begin{equation} \label{eq:annih op}
    \mathcal{O}^{X}_\Delta(z,\bar{z}) = \int_{0}^\infty \frac{d\omega}{\omega} \omega^{\Delta} ~ a_{X}(\omega,z,\bar{z})\hspace{2pt}.
\end{equation}
One can verify that $\mathcal{O}^{X}_\Delta(z,\bar{z})$ transforms under the 4D Lorentz group as a 2D conformal primary operator with weight $\Delta$ and spin $s_X$ (equal to the bulk helicity) located at the point $(z,\bar{z})$ on the celestial sphere. Incoming operators may also be constructed but in this paper  we focus on the outgoing case.

\subsubsection {Celestial OPEs}

In the derivation of soft theorems and symmetry algebras, it  is very useful to study holomorphic OPEs between operators, $\mathcal{O}^X_{\Delta_X}(z,\bar{z})\mathcal{O}^Y_{\Delta_Y}(z',\bar{z}')$, which involve taking $z - z' \rightarrow 0$ while holding $\bar{z} - \bar{z}\hspace{1pt}'$ fixed.  We may equivalently obtain  the independence  of $z$ and $\bar{z}$  either by permitting complex Minkowski  momenta or rotating to signature $(2,2)$ Klein space with real momenta. In either case the Lorentz group effectively becomes $S L_{2}(\mathbb{R})\times \overline{ S L}_{2}(\mathbb{R})$. Here we focus  on the holomorphic limit, but of course there is a conjugate anti-holomorphic one.

Holomorphic OPEs are obtained by Mellin transforms of the collinear expansions of scattering amplitudes. The singular part of the holomorphic OPE is given by \cite{Pate:2019lpp,Himwich:2021dau}\footnote{We omit any `multi-particle operators' which may appear in the celestial OPE \cite{Ball:2023sdz, Guevara:2024ixn}.}
\begin{align}
        \mathcal{O}^X_{\Delta_X}(z,\bar{z}) \mathcal{O}^Y_{\Delta_Y}(0,0) \sim \frac{1}{z} \sum_{Z} &\gamma_{XYZ} \sum_{m = 0}^\infty \frac{1}{m!} B(\Delta_X + s_Y - s_Z -1 +m, \Delta_Y + s_X - s_Z - 1) \nonumber \\
        & \times \bar{z}^{s_X + s_Y - s_Z - 1+m}\hspace{4pt} \bar{\partial}^m \mathcal{O}^{Z}_{\Delta_X + \Delta_Y + s_X + s_Y - s_Z - 2}(0,0)\hspace{2pt},
    \label{eqn: general OPE expression}
\end{align}
where $B(a,b) = \Gamma(a) \Gamma(b)/\Gamma(a+b)$ is the Euler beta function, $Z$ sums over conformal primaries with $s_Z < s_X + s_Y$, and $\gamma_{XYZ}$ is a proportionality constant appearing in the collinear factorization.

\subsubsection {Symmetry Algebras} \label{sec:review_symalg}

Celestial OPEs are useful because they elucidate symmetries of the theory which are hard to find  from the bulk perspective. 

To see this, one defines the conformally-soft operators 
\begin{equation} \label{eq:defn_conf_soft}
    \mathcal{O}_{\text{CS}}^{X,k}(z,\bar z) \equiv \lim_{\varepsilon \hspace{1pt} \rightarrow \hspace{1pt} 0} \varepsilon \hspace{2pt} \mathcal{O}_{k + \varepsilon}^{X}(z,\bar z)\hspace{2pt},
\end{equation}
where $k \in \mathbb{Z}$ for bosonic operators and $k \in \mathbb{Z} + \tfrac{1}{2}$ for fermionic operators. The RHS is not trivial because the poles in the OPE can cancel the zero for (half) integer $k\le s_X$. In the holomorphic expansion, we may then consistently  expand in $\bar{z}$
\begin{equation}
\mathcal{O}_{\text{CS}}^{X,k}(z,\bar{z}) = \sum_{n = \frac{1}{2}(k-s_X)}^{-\frac{1}{2}(k-s_X)} \mathcal{O}_{\text{CS},n}^{X,k}(z) \hspace{2pt} \bar{z}^{\hspace{1pt}-n-\frac{1}{2}(k-s_X)}\hspace{2pt},
    \label{eqn: mode expansion}
\end{equation}
where $n$ denotes the $\bar{z}$ mode number.  This mode expansion contains a finite number of terms,\footnote{Terms outside the given range have trivial holomorphic OPEs due to the $\varepsilon $ rescaling and may be set to zero in the holomorphic expansion.} and each such conformally soft operator is thereby decomposed as a chiral current transforming in a negative weight ($s_X-k+1$)-dimensional representation of  $\overline{S L}_{2}(\mathbb{R})$. To organize the resulting {\it chiral soft algebra}, it is convenient to redefine the currents according to \cite{strominger2021w}
\begin{equation} \label{defn_hat_rescaling}
    \mathcal{R}^{X,p}_{n} \equiv \Gamma(p+n)\Gamma(p-n) \mathcal{O}_{\text{CS},n}^{X,2-2p + s_X} \ , \qquad  p = 1, \frac{3}{2},2,\frac{5}{2},...
\end{equation}
The restriction on the sum \eqref{eqn: mode expansion} means that all modes are defined in the wedge $1-p \leq n \leq p-1$ where $n$ is an integer (respectively half-integer) when $p$ is an integer (respectively half-integer). $\mathcal{R}^{X,p}_{n}(z)$ has holomorphic weight $h = 1-p + s_X$. 

Further expanding in $z$ (keeping all powers)
\begin{equation}
   {\mathcal{R}}^{X,p}_{n}(z) = \sum_{m} z^{-m + p - 1 - s_X} {\mathcal{R}}^{X,p}_{n,m} \ , \qquad {\mathcal{R}}^{X,p}_{n,m} = \oint_0 \frac{dz}{2\pi i} \hspace{2pt} z^{m - p + s_X} {\mathcal{R}}^{X,p}_{n}(z) \hspace{2pt},
   \label{eqn: m mode expansion}
\end{equation}
we may define  \cite{guevara2021holographic, Himwich:2021dau}
\begin{equation}
\Big[{\mathcal{R}}^{X,p}_{n,m},\mathcal{R}^{Y,p'}_{n',m'}\Big] = \oint_0 \frac{dz_2}{2\pi i} \hspace{2pt} z_2^{m' - p' + s_Y} \oint_{z_2} \frac{dz_1}{2\pi i} \hspace{2pt} z_1^{m -p + s_X} \hspace{3pt} {\mathcal{R}}^{X,p}_{n}(z_1) \hspace{2pt} {\mathcal{R}}^{Y,p'}_{n'}(z_2) \hspace{2pt}.
\end{equation}
Explicit commutators $\big[{\mathcal{R}}^{X,p}_{n,m},{\mathcal{R}}^{Y,p'}_{n',m'}\big]$ may then be evaluated starting from the OPE \eqref{eqn: general OPE expression}, as we shall see below. 

Wilsonian corrections to chiral soft algebras arise only from poles in the OPE and hence are very highly constrained by nonrenormalization theorems. This enables an exact calculation of the tree-level chiral soft algebra in the full Wilsonian theory at any point in ${\mathcal{M}}_\phi$. In contrast the soft theorems controlling the behavior of single insertions of soft currents involve the full OPE. These can receive a finite number of corrections which  grows with the spin of the current, and hence for sufficiently high spin requires the full Wilsonian action \cite{Pate:2019lpp, Himwich:2021dau}.

\section {Chiral Soft Algebras on $\mathcal{M}_{\phi}$ } \label{sec:holomorphicOPEs_woSings}

In this section, we construct  the chiral soft algebra of $\mathcal{N} = 2$ effective field theories with action \eqref{eq:Leff_component_fields} at generic, smooth points in the moduli space $\mathcal{M}_{\phi}$, and study its properties. 
\subsection{Bulk Supercharges}
We begin by expanding the scalar field $\phi = a + \varphi$ as in \eqref{eq: scalar field vev expansion} -- the $a$ coordinate is a vev that parametrizes one's location in $\mathcal{B}$ while $\varphi$ is the dynamical field that creates single-particle scattering states. 
Keeping terms up to cubic order in the dynamical fields, the effective Lagrangian reads
\begin{equation}
   \begin{split}
       \mathcal{L}_{\text{eff}} &= - \frac{\text{Im} \hspace{2pt}\tau(a)}{4\pi} \bigg(\frac{1}{4}F^{\mu \nu} F_{\mu \nu} + \partial_\mu \varphi \partial^\mu \overline{\varphi} + i \psi_I \sigma^\mu \partial_\mu \bar{\psi}^I  \bigg) +  \frac{\text{Re}\hspace{2pt}\tau(a)}{16 \pi}F^{\mu \nu} \widetilde{F}_{\mu \nu}\\
       &\hspace{30pt} - \frac{1}{8\pi i} \big(\tau'(a) \varphi - \bar{\tau}'(a)\overline{\varphi}\big)\bigg(\frac{1}{4}F^{\mu \nu} F_{\mu \nu} + \partial_\mu \varphi \partial^\mu \overline{\varphi} + i \psi_I \sigma^\mu \partial_\mu \bar{\psi}^I  \bigg) \\
       &\hspace{30pt}+\frac{\sqrt{2}}{32\pi i} \bigg(\tau'(a) \varepsilon^{IJ} \psi_I \sigma^{\mu \nu} \psi_J F_{\mu \nu} + \bar{\tau}'(a) \varepsilon^{IJ} \bar{\psi}_I \bar{\sigma}^{\mu \nu} \bar{\psi_J} F_{\mu \nu}\bigg) \\
       &\hspace{30pt} + \frac{1}{32\pi} \big(\tau'(a) \varphi + \bar{\tau}'(a)\overline{\varphi}\big) F^{\mu \nu}\tilde{F}_{\mu \nu} + \cdots
       \label{eq: cubic Lagrangian}
   \end{split}
\end{equation}
The $\mathcal{N} = 2$ supersymmetry is enacted by two pairs of supercharges labelled $Q_{I\alpha}$ and $\overline{Q}_{I\dot{\alpha}}$ which are two-component Weyl spinors obeying the Hermiticity condition $(Q_{I\alpha})^\dagger = \overline{Q}^{I}_{\dot{\alpha}} = \varepsilon^{IJ} \hspace{1pt} \overline{Q}_{J\dot \alpha}.$ They have the following action on the bulk spacetime fields \cite{labastida2005topological} 
\begin{equation}
    \begin{split}
        &\Big\llbracket Q_{I\alpha}, \varphi\Big\rrbracket = -i\sqrt{2} \hspace{2pt} \psi_{I\alpha} \hspace{80pt} \Big\llbracket \overline{Q}_{I \dot \alpha}, \varphi\Big\rrbracket = 0\\
        &\Big\llbracket Q_{I\alpha}, \overline{\varphi}\Big\rrbracket = 0 \hspace{120.5pt} \Big\llbracket \overline{Q}_{I \dot \alpha}, \overline{\varphi}\Big\rrbracket = -i \sqrt{2} \hspace{2pt} \overline{\psi}_{I\dot{\alpha}}\\
        &\Big\llbracket Q_{I\alpha}, \psi_{J\beta}\Big\rrbracket = i\varepsilon_{IJ} \sigma^{\mu \nu}_{\alpha \beta} F_{\mu \nu} \hspace{61pt} \Big\llbracket \overline{Q}_{I\dot \alpha}, \psi_{J\beta}\Big\rrbracket = \sqrt{2} \varepsilon_{IJ} \sigma^{\mu}_{\beta \dot{\alpha}} \partial_\mu \varphi\\ 
        &\Big\llbracket Q_{I\alpha}, \overline{\psi}_{J\dot \beta}\Big\rrbracket = -\sqrt{2} \varepsilon_{IJ} \sigma^\mu_{\alpha \dot{\beta}} \partial_\mu \overline{\varphi} \hspace{40pt} \Big\llbracket \overline{Q}_{I\dot \alpha}, \overline{\psi}_{J\dot \beta}\Big\rrbracket = -i\varepsilon_{IJ} \overline{\sigma}^{\mu \nu}_{\dot{\alpha} \dot{\beta}} F_{\mu \nu}\\
        &\Big\llbracket Q_{I\alpha}, A^\mu\Big\rrbracket = \sigma^\mu_{\alpha \dot{\alpha}} \hspace{2pt} \bar{\psi}_I^{\dot{\alpha}} \hspace{86pt} \Big\llbracket \overline{Q}_{I \dot \alpha}, A^\mu \Big\rrbracket =  -\psi^\alpha_I \sigma^\mu_{\alpha \dot{\alpha}} \ , \\
    \end{split}\label{ssy}
\end{equation}
where we have dropped terms proportional to the Lagrange multiplier fields. The brackets $\big\llbracket \cdot,\cdot \big\rrbracket$ are the usual bulk (anti-)commutators,  not to be confused with the $\big[\cdot,\cdot\big]$ boundary commutators defined via contour integration.

\begin{figure}
   \centering
    \resizebox{0.65\textwidth}{!}    {
        \input{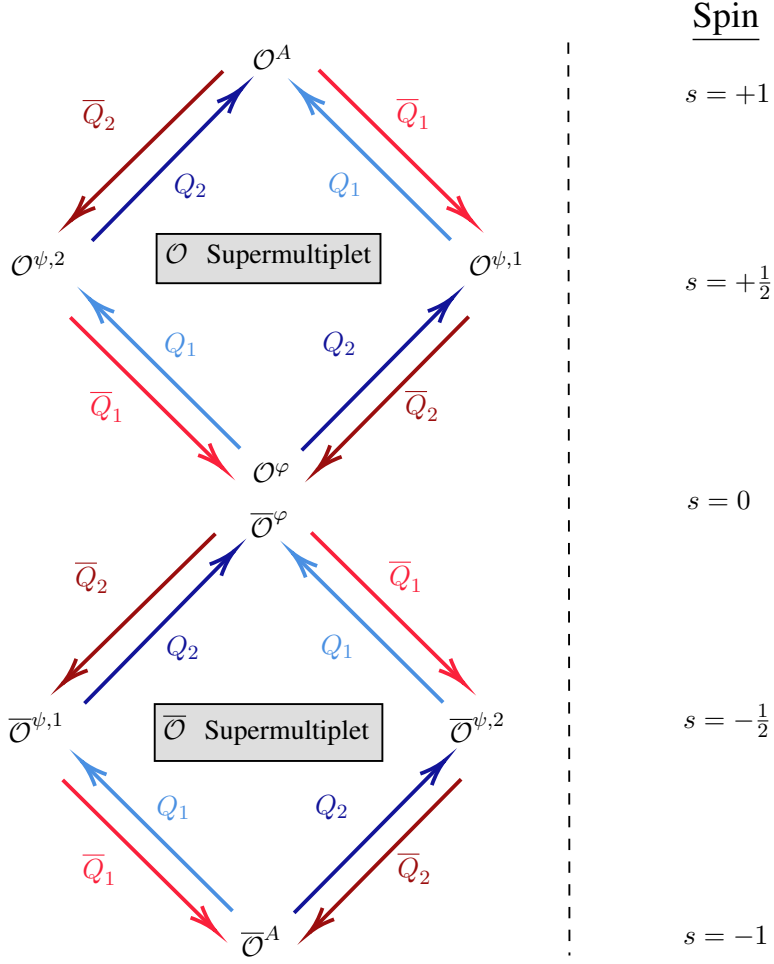}
    }
    \caption{How various spacetime supercharges act on conformal primary operators in the CCFT.}
    \label{fig:SUSY Modes}
\end{figure}

The 2D conformal primary operators live in a vector multiplet under bulk $\mathcal{N} = 2$ supersymmetry (see Appendix \ref{appendix: conventions} for conventions).
\begin{equation}
    \text{$\mathcal{N} = 2$ Vector Multiplet}: \hspace{15pt}\begin{cases}
        \mathcal{O}^{A}_{\Delta}(z,\bar z) \hspace{10pt},\hspace{10pt} \mathcal{O}^{\psi,I}_{\Delta}(z,\bar z) \hspace{10pt},\hspace{10pt}  \mathcal{O}^{\varphi}_{\Delta}(z,\bar z) \\
        \overline{\mathcal{O}}^{A}_{\Delta}(z,\bar z) \hspace{10pt},\hspace{10pt}  \overline{\mathcal{O}}^{\psi,I}_{\Delta}(z,\bar z) \hspace{10pt},\hspace{10pt}  \overline{\mathcal{O}}^{\varphi}_{\Delta}(z,\bar z).
    \end{cases}
\end{equation}
 For simplicity, here we consider only operators corresponding to outgoing particles: our conventions relating them to the Mellin transform of  annihilation operators are given in Appendix \ref{appendix: conventions}. The barred and unbarred multiplets each transform as irreducible representations under bulk supersymmetry (related by CPT conjugation). These relations are summarized in Figure \ref{fig:SUSY Modes}. Supersymmetry acts on the unbarred multiplet as:
\begin{equation} \label{eq:SUSY action_plus}
    \begin{split}
        &\Big\llbracket Q_{I\alpha}\hspace{2pt},\hspace{2pt}\mathcal{O}^{\varphi}_\Delta(z,\bar{z})\Big\rrbracket = -i\sqrt{2} \hspace{2pt} \varepsilon_{IJ}  |z\rangle_\alpha \mathcal{O}^{\psi,J}_{\Delta + 1/2} \hspace{39pt} \Big\llbracket \hspace{2pt}\overline{Q}_{I\dot{\alpha}}\hspace{2pt},\hspace{2pt}\mathcal{O}^{\varphi}_\Delta(z,\bar{z})\Big\rrbracket = 0\\
        &\Big\llbracket Q_{I\alpha}\hspace{2pt},\mathcal{O}^{\psi,J}_{\Delta}(z,\bar{z})\Big\rrbracket = -\sqrt{2} \hspace{2pt} \delta^J_I |z\rangle_{\alpha} \mathcal{O}^{A}_{\Delta + 1/2} \hspace{42pt} \Big\llbracket\hspace{2pt}\overline{Q}_{I\dot{\alpha}}\hspace{2pt},\mathcal{O}^{\psi,J}_{\Delta}(z,\bar{z})\Big\rrbracket = - i \sqrt{2} \hspace{2pt} \delta^J_I [\bar{z}|_{\dot{\alpha}}  \mathcal{O}^{\varphi}_{\Delta + 1/2}\\
        &\Big\llbracket Q_{I\alpha}\hspace{2pt},\hspace{1pt}\mathcal{O}^{A}_\Delta(z,\bar{z})\Big\rrbracket = 0 \hspace{133pt} \Big\llbracket \hspace{2pt}\overline{Q}_{I\dot{\alpha}}\hspace{2pt},\hspace{1pt}\mathcal{O}^{A}_\Delta(z,\bar{z})\Big\rrbracket = \sqrt{2} \hspace{2pt} \varepsilon_{IJ} [\bar{z}|_{\dot{\alpha}} \mathcal{O}^{\psi,J}_{\Delta + 1/2}\ , \\
    \end{split}
\end{equation}
where, as usual, the spinors are defined as $|z\rangle_\alpha \equiv (1,z)$ and $|\bar{z}]_{\dot{\alpha}} \equiv (1,\bar{z})$. The barred multiplet transforms as:
\begin{equation} \label{eq:SUSY action_minus}
    \begin{split}
        &\Big\llbracket Q_{I\alpha}\hspace{2pt},\hspace{2pt}\overline{\mathcal{O}}^{\varphi}_\Delta(z,\bar{z})\Big\rrbracket = 0 \hspace{133pt} \Big\llbracket\hspace{2pt}\overline{Q}_{I\dot{\alpha}}\hspace{2pt},\hspace{2pt}\overline{\mathcal{O}}^{\varphi}_\Delta(z,\bar{z})\Big\rrbracket = -i \sqrt{2} \hspace{2pt} \varepsilon_{IJ} [\bar{z}|_{\dot{\alpha}} \overline{\mathcal{O}}^{\psi,J}_{\Delta + 1/2}\\
        &\Big\llbracket Q_{I\alpha}\hspace{2pt},\overline{\mathcal{O}}^{\psi,J}_\Delta(z,\bar{z})\Big\rrbracket = i \sqrt{2} \hspace{2pt} \delta^J_I |z\rangle_{\alpha} \overline{\mathcal{O}}^{\varphi}_{\Delta + 1/2} \hspace{47.5pt} \Big\llbracket \hspace{2pt}\overline{Q}_{I\dot{\alpha}}\hspace{2pt},\overline{\mathcal{O}}^{\psi,J}_\Delta(z,\bar{z})\Big\rrbracket = -\sqrt{2} \hspace{2pt} \delta^J_I [\bar{z}|_{\dot{\alpha}}  \overline{\mathcal{O}}^{A}_{\Delta + 1/2}\\
        &\Big\llbracket Q_{I\alpha}\hspace{2pt},\hspace{1pt}\overline{\mathcal{O}}^{A}_\Delta(z,\bar{z})\Big\rrbracket = \sqrt{2} \hspace{2pt} \varepsilon_{IJ} |z\rangle_\alpha \overline{\mathcal{O}}^{\psi,J}_{\Delta + 1/2} \hspace{53pt} \Big\llbracket \hspace{2pt}\overline{Q}_{I\dot{\alpha}}\hspace{2pt},\hspace{1pt}\overline{\mathcal{O}}^{A}_\Delta(z,\bar{z})\Big\rrbracket = 0 \ .
    \end{split}
\end{equation}
Note that these are not the usual 2D  supersymmetry transformations, which involve differentiation with respect to $z$ or $\bar z$ (rather than multiplication by them). Related results for bulk supersymmetry charges acting on CCFT operators  for different examples can be found in \cite{jiang2022holographic, Fotopoulos:2020bqj, Ball:2023qim}.

\subsection {Chiral Boundary OPEs}

OPEs in the CCFT may be determined from three-particle scattering amplitudes or splitting functions. In our case, they are  
\begin{equation}
    \begin{split}
        \mathcal{O}_{\Delta}^{A}(z,\bar{z}) \mathcal{O}_{\Delta'}^{A}(0,0) &\sim \frac{1}{z}\left(-\frac{\mu}{2}\right) \sum_{m = 0}^\infty \frac{1}{m!} B(\Delta + m,\Delta') \bar{z}^{1+m}\hspace{2pt}\bar{\partial}^m \overline{\mathcal{O}}^{\varphi}_{\Delta + \Delta'}(0,0)\hspace{2pt}, \\
        \mathcal{O}_{\Delta}^{A}(z,\bar{z}) \mathcal{O}_{\Delta'}^{\varphi}(0,0) &\sim \frac{1}{z}\left(-\frac{\mu}{2}\right) \sum_{m = 0}^\infty \frac{1}{m!} B(\Delta + m,\Delta' + 1) \bar{z}^{1+m}\hspace{2pt}\bar{\partial}^m \overline{\mathcal{O}}^{A}_{\Delta + \Delta'}(0,0)\hspace{2pt}, \\
        \mathcal{O}_{\Delta}^{\psi,I}(z,\bar{z}) \mathcal{O}_{\Delta'}^{\psi,J}(0,0) &\sim \frac{i}{z}\left(-\frac{\mu}{2}\right) \varepsilon^{IJ}\sum_{m = 0}^\infty \frac{1}{m!} B(\Delta + m + \tfrac{1}{2},\Delta' + \tfrac{1}{2}) \bar{z}^{1+m}\hspace{2pt}\bar{\partial}^m \overline{\mathcal{O}}^{A}_{\Delta + \Delta'}(0,0) \hspace{2pt},\\ 
        \mathcal{O}_{\Delta}^{\psi,I}(z,\bar{z}) \mathcal{O}_{\Delta'}^{A}(0,0) &\sim \frac{i}{z}\left(-\frac{\mu}{2}\right) \sum_{m = 0}^\infty \frac{1}{m!} B(\Delta + m + \tfrac{1}{2},\Delta') \bar{z}^{1+m}\hspace{2pt}\bar{\partial}^m \overline{\mathcal{O}}^{\psi,I}_{\Delta + \Delta'}(0,0) \hspace{2pt},
        \label{eq: holomorphic OPEs}
    \end{split}
\end{equation}
where we have defined the coupling constant
\begin{equation} \label{eq:mubar}
    \mu \equiv  \frac{i}{16\pi }\frac{\overline{\tau}'(a)}{(\text{Im}\hspace{2pt} \tau(a)/4\pi)^{3/2}}\hspace{2pt}.
\end{equation}
Subleading terms arise from Wilsonian corrections but are nonsingular for $z \to 0$ at fixed $\bar z$.

\subsection {Chiral Soft Algebras}
\label{sec: Holomorphic soft algebras}
The OPEs \eqref{eq: holomorphic OPEs} have poles at certain (half) integral values of $\Delta \le 1$.  Rescaled generators of the chiral soft algebra with finite OPEs are defined at these poles as in \eqref{defn_hat_rescaling} and \eqref{eqn: m mode expansion}: 
\begin{equation} \label{eq: N2 vector multiplet}
    \text{$\mathcal{N} = 2$ Vector Multiplet}: \hspace{15pt}\begin{cases} \hspace{5pt}
        \mathcal{R}^{A,p}_{n,m} \hspace{10pt},\hspace{10pt} \mathcal{R}^{\psi,I,p}_{n,m} \hspace{10pt},\hspace{10pt}  \mathcal{R}^{\varphi,p}_{n,m} \\
        \hspace{5pt} \overline{\mathcal{R}}{}^{A,p}_{n,m} \hspace{10pt},\hspace{10pt} \overline{\mathcal{R}}{}^{\psi,I,p}_{n,m} \hspace{10pt},\hspace{10pt}  \overline{\mathcal{R}}{}^{\varphi,p}_{n,m} \hspace{2pt} ,\\
    \end{cases}
\end{equation}
for $p = 1,\frac{3}{2},2, \cdots$ and the wedge $1-p \leq n \leq p-1$.  

These operators generate infinitesimal (in general spontaneously broken) symmetries on the quantum states of the celestial CFT on a circle. For example, the $\mathcal{R}{}^{A,1}_{0,m}$ generate large gauge symmetries, while  ${\mathcal{R}}^{\varphi,1}_{0,0}$ generates motion on the moduli space, following from the geometric soft theorem \cite{Kapec:2022axw, Derda:2024jvo, Cheung:2021yog, Cheung:2022vnd}
\be \overline{\mathcal{R}}{}^{ \varphi,1}_{0,0}|a, \bar a \rangle=\p^{\bar a}|a, \bar a\rangle = \frac{4\pi}{{\rm Im} \hspace{2pt} \tau} \p_{a}|a, \bar a\rangle,\ee 
where $|a , \bar a \rangle$ is the 2D state associated to the CFT at a point where the scalar $\phi$  has vev $a$. Note however that shifts of $\bar a$ are not generated by any element of this chiral algebra: rather they appear in the antichiral algebra arising from the expansion around  $\bar z \to 0$ with fixed $z$. Hence, on its own $\overline{\mathcal{R}}{}^{ \varphi,1}_{0,0}$ generates an infinitesimal transformation off of the physical moduli space in which the vevs of $\phi$ and $\bar \phi$ are complex conjugates.\footnote{The same can be said of the SUSY variation by the nonhermitian supercharges appearing in \eqref{ssy} which vary $\phi$ but not $\bar \phi$. Interestingly in Klein space they are independent and the transformations are generated  by integrals of the chiral currents over null lines of fixed $z$ or $\bar z$, see below. A cogent analysis of transport on the moduli space using both chiral and antichiral soft modes can be found in \cite{Kapec:2022axw}}

\subsubsection{Commutators}
Commutators of these operators are determined by taking contour integrals of the conformally soft limit of OPEs in Equation \eqref{eq: holomorphic OPEs} and rescaling the generators as in Equation \eqref{defn_hat_rescaling}\footnote{This generalizes the  commutators between photons and scalars found in  \cite{Mago:2021wje,Ren:2022sws,Melton:2022fsf} to the supersymmetric case.} \begin{equation}
\begin{split}
    \left[\mathcal{R}^{A, p}_{n,m} \ , \mathcal{R}^{A,p'}_{n',m'}\right]&= \mu \left(n(p'-1) - n'(p-1)\right) \overline{\mathcal{R}}{}_{n+n',m+m'}^{\varphi, p+p'-2}\hspace{2pt},\\
    \left[\mathcal{R}_{n,m}^{A,p} \ , \mathcal{R}_{n',m'}^{\varphi,p'}\right]&= \mu \left(n(p'-1) - n'(p-1)\right) \overline{\mathcal{R}}{}_{n+n',m+m'}^{A,p+p'-2}\hspace{2pt},\\
    \left[\mathcal{R}_{n,m}^{\psi,I,p} \ , \mathcal{R}_{n',m'}^{\psi,J,p'}\right]&= i  \mu \left(n(p'-1) - n'(p-1)\right) \varepsilon^{IJ} \hspace{2pt} \overline{\mathcal{R}}{}_{n+n',m+m'}^{A,p+p'-2}\hspace{2pt},\\
    \left[\mathcal{R}_{n,m}^{\psi,I,p} \ , \mathcal{R}_{n',m'}^{A,p'}\right]&= i \mu
\left(n(p'-1) - n'(p-1)\right) \overline{\mathcal{R}}{}_{n+n',m+m'}^{\psi,I,p+p'-2}\hspace{2pt}.
\label{eq: soft algebra}
\end{split}
\end{equation}
These commutators are uncorrected by higher Wilsonian corrections because, as note above, such corrections do not contribute to poles in the OPE.\footnote{There are a few terms such as $F^3$ which can affect the pole but are forbidden by supersymmetry \cite{Ball:2023qim}.} They indicate a rich interplay between the moduli space geometry and the chiral soft algebra.  The first equation implies that the commutator of two $p={3 \over 2}$ gauge generators gives a shift in the moduli space. Hence the total infinite-dimensional space of vacua  is not simply a product of the degenerate scalar and large gauge vacua.\footnote{A similar conclusion may be deduced in the case of Einstein-Maxwell theory where the commutator of two gauge generators produces a BMS translation.} The second equation states that the positive and negative chirality soft photons rotate into one another under motion in the scalar moduli space. Note also that the  barred generators are in the center of the chiral  soft algebra. (In the anti-chiral  algebra, the unbarred generators are central).  There is a closed subalgebra with the restricted values of $p=1,\frac{3}{2},2.$ We leave exploration of  this interesting structure to future work.


\subsubsection {SL$_2(\mathbb{Z})$ Covariance}
\label{sec: SL2 covariance}
This subsection analyzes  SL$_2(\mathbb{Z})$ covariance of the soft algebra.
A function $f: \mathcal{B} \rightarrow \mathbb{C}$ has  $SL_2(\mathbb{Z})$ weight $(N,M)$ if, under an $SL_2(\mathbb{Z})$ transformation \eqref{eqn: SL(2,Z) transform} taking $(a, a_D) \mapsto (a', a_D')$, the function transforms as
\begin{equation}
    f(a_D',a') = (y \tau(a) + z)^N (y \bar{\tau}(a) + z)^M f(a_D,a)\hspace{2pt}.
\end{equation}
One then finds the $SL_2(\mathbb{Z})$ weights \begin{equation}
    \frac{d\tau(a)}{da}: \big(-3,0\big) \ , \qquad \frac{d\bar{\tau}(a)}{da}: \big(0,-3\big)\ , \qquad \text{Im} \hspace{2pt} \tau : \big(-1,-1\big)\ , \qquad (\text{Im} \hspace{2pt} \tau)^{-3/2}: \big(\tfrac{3}{2},\tfrac{3}{2}\big)\hspace{2pt},
\end{equation}
\begin{equation}\label{eq:mubar SL2Z weight}
    \mu : \big(\tfrac{3}{2},-\tfrac{3}{2}\big)\ , \hspace{22pt} \overline{\mu}: \big(-\tfrac{3}{2},\tfrac{3}{2}\big)\hspace{2pt}.
\end{equation}
To linear order in the field fluctuations, the self-dual field strength  $F^+_{\mu\nu},\psi_I$, and $\varphi$ transform homogeneously  with weights \begin{align}
    F^{+}_{\mu \nu}:\big(0,1\big)\ , \hspace{40pt} \psi_I: \big(1,0\big)\ , \hspace{40pt} \varphi: \big(1,0\big)\hspace{2pt}.
\end{align}
On the other hand, $F^-_{\mu \nu},\bar{\psi}^I$ and $\bar{\varphi}$ -- being related by Hermitian conjugation -- have the opposite weights. The celestial operators constructed from these fields in Appendix \ref{appendix: conventions} have weights 
\begin{equation}
    \begin{split}
    \mathcal{O}^{X}_{\Delta}(z,\bar z) ~ ~ \text{and} ~ ~ \mathcal{R}^{X,p}_{n,m} : \big(\tfrac{1}{2},-\tfrac{1}{2}\big) \ , \hspace{30pt} \overline{\mathcal{R}}{}^{X}_{\Delta}(z,\bar z) ~ ~ \text{and} ~ ~ \overline{\mathcal{O}}^{X,p}_{n,m}: \big(-\tfrac{1}{2},\tfrac{1}{2}\big) \ ,
    \label{eq: SL2 weights}
    \end{split}
\end{equation}
for $X \in \{A,\psi_I, \varphi\}.$

We are now ready to verify $SL_2(\mathbb{Z})$ covariance of the commutation relations. For example, consider the first commutator in Equation \eqref{eq: soft algebra}
\begin{equation}
    \left[\mathcal{R}_{n,m}^{A,p} \ , \mathcal{R}_{n',m'}^{A,p'}\right]= \mu \left(n(p'-1) - n'(p-1)\right) \overline{\mathcal{R}}{}_{n+n',m+m'}^{\varphi,p+p'-2}\hspace{2pt},
\end{equation}
which is computed in some particular $SL_2(\mathbb{Z})$ frame.  From Equation \eqref{eq:mubar SL2Z weight} and Equation \eqref{eq: SL2 weights}, we have
\begin{equation}
    \begin{split}
        \mu' &= (c \tau + d)^{3/2}(c \bar \tau + d)^{-3/2} \hspace{2pt} \mu \hspace{2pt},\\
        \big(\mathcal{R}^{A,p}_{n,m}\big)' &= (c \tau + d)^{1/2}(c \bar \tau + d)^{-1/2} \hspace{2pt}\mathcal{R}^{A,p}_{n,m}\hspace{2pt},\\
        \big(\overline{\mathcal{R}}{}^{\varphi,p}_{n,m}\big)' &= (c \tau + d)^{-1/2}(c \bar \tau + d)^{1/2} \hspace{2pt}\overline{\mathcal{R}}{}^{\varphi,p}_{n,m}\hspace{2pt}.
    \end{split}
\end{equation}
It follows that both sides of the equation transform with  weight $(1,-1)$, which therefore is valid in every $SL_2(\mathbb{Z})$ frame.  This, in particular, guarantees that the chiral algebra is well-defined in the neighborhood of strongly coupled singular points with $SL_2(\mathbb{Z})$ monodromies. 

\subsubsection{Supersymmetry of the Chiral Algebra}
\label{sec: bulk SUSY}

Using equations \eqref{defn_hat_rescaling}, \eqref{eqn: m mode expansion}, and \eqref{eq:SUSY action_plus}, one can determine how $\mathcal{N} = 2$ supersymmetry acts on the generators of the holomorphic symmetry algebra. In the unbarred multiplet
\begin{equation}
    \begin{split}
        &\Big\llbracket Q_{I\alpha}, \mathcal{R}^{\varphi,p}_{n,m}\Big\rrbracket = -i \sqrt{2} \hspace{2pt} \varepsilon_{IJ} \hspace{2pt} \mathcal{R}^{\psi,J,p}_{n,m + \alpha}\hspace{33.5pt} \Big\llbracket \overline{Q}_{I\dot\alpha}, \mathcal{R}^{\varphi,p}_{n,m}\Big\rrbracket = 0\\
        &\Big\llbracket Q_{I\alpha},\mathcal{R}^{\psi,J,p}_{n,m}\Big\rrbracket = -\sqrt{2} \hspace{2pt} \delta^J_I \hspace{2pt} \mathcal{R}^{A,p}_{n,m + \alpha} \hspace{37pt} \Big\llbracket \overline{Q}_{I\dot\alpha}, \mathcal{R}^{\psi,J,p}_{n,m}\Big\rrbracket = -i \sqrt{2} \hspace{2pt} \delta_I^J \hspace{2pt} (p -2 \dot{\alpha} n-1)\mathcal{R}^{\varphi,p-1/2}_{n + \dot{\alpha},m} \\
        &\Big\llbracket Q_{I\alpha},\mathcal{R}^{A,p}_{n,m}\Big\rrbracket = 0\hspace{111pt} \Big\llbracket \overline{Q}_{I\dot\alpha},\mathcal{R}^{A,p}_{n,m}\Big\rrbracket = \sqrt{2} \hspace{2pt} \varepsilon_{IJ}\hspace{2pt} (p - 2\dot{\alpha} n-1)\mathcal{R}^{\psi,J,p-1/2}_{n + \dot{\alpha},m} \ ,\\
    \end{split}
\end{equation}
where $\alpha, \dot{\alpha} = \pm \frac{1}{2}$. Similarly, for the barred multiplet
\begin{equation}
    \begin{split}
        &\Big\llbracket Q_{I\alpha}, \overline{\mathcal{R}}{}^{\varphi,p}_{n,m}\Big\rrbracket = 0 \hspace{109pt} \Big\llbracket \overline{Q}_{I\dot\alpha}, \overline{\mathcal{R}}{}^{\varphi,p}_{n,m}\Big\rrbracket = -i \sqrt{2}\hspace{2pt} \varepsilon_{IJ} \hspace{2pt} (p -2 \dot{\alpha} n-1)\overline{\mathcal{R}}{}^{\psi,J,p-1/2}_{n + \dot{\alpha},m}\\
        &\Big\llbracket Q_{I\alpha}, \overline{\mathcal{R}}{}^{\psi,J,p}_{n,m}\Big\rrbracket = i \sqrt{2} \hspace{2pt} \delta_I^J \hspace{2pt} \overline{\mathcal{R}}{}^{\varphi,p}_{n,m + \alpha}  \hspace{40pt} \Big\llbracket \overline{Q}_{I\dot\alpha}, \overline{\mathcal{R}}{}^{\psi,J,p}_{n,m}\Big\rrbracket = -\sqrt{2} \hspace{2pt} \delta^{J}_I \hspace{2pt} (p - 2\dot{\alpha} n-1)\overline{\mathcal{R}}{}^{A,p-1/2}_{n + \dot{\alpha},m}\\
        &\Big\llbracket Q_{I\alpha}, \overline{\mathcal{R}}{}^{A,p}_{n,m}\Big\rrbracket = \sqrt{2} \hspace{2pt} \varepsilon_{IJ} \hspace{2pt} \overline{\mathcal{R}}{}^{\psi,J,p}_{n,m + \alpha}\hspace{44.5pt} \Big\llbracket \overline{Q}_{I\dot\alpha}, \overline{\mathcal{R}}{}^{A,p}_{n,m}\Big\rrbracket = 0 \ .\\
    \end{split}
\end{equation}
We observe that $Q_{I\alpha}$ acts on the holomorphic $m$ index while $\overline{Q}_{I \dot{\alpha}}$ acts on the anti-holomorphic $n$ index while giving an additional prefactor which vanishes when the operators on the right hand side are taken outside the wedge $p - 1 \leq n \leq 1-p.$

One may check that the Jacobi identity is obeyed for mixed bulk and boundary commutators. This implies that supersymmetry is an outer derivation of the soft algebra. In supergravity this global supersymmetry is presumably promoted to a generator of the chiral algebra \cite{Fotopoulos:2020bqj, Tropper:2024kxy}.

\subsection{Hypermultiplets}

Abelian $\mathcal{N} = 2$ gauge theories may also contain a bulk hypermultiplet of charged particles with $n_e \in \mathbb{Z}$ units of electric charge and $n_m \in \mathbb{Z}$ units of magnetic charge. This $\mathcal{N} = 2$ hypermultiplet is comprised of an $SU(2)_R$ doublet of complex scalar fields, $q_I$, satisfying the Hermiticity condition $q_I^\dagger = \bar{q}^I$ as well as two singlet Weyl spinors, $\lambda$ and $\chi$.
The action for the full theory is simply
\begin{equation}
    \mathcal{L} = \mathcal{L}_{\text{vector}} + \mathcal{L}_{\text{hyper}}.
\end{equation}
Relevant  details of  $\mathcal{L}_{\text{hyper}}$ are given in Appendix \ref{appendix: hypermultiplet}. Importantly, all three-point amplitudes in $\mathcal{L}_{\text{hyper}}$ involve one particle in the vector multiplet and two particles in the hypermultiplet due to charge conservation. Holomorphic OPEs and symmetry algebras are all proportional to the coupling constant $n_e + \bar \tau n_m$ appearing in these three-point amplitudes.  Hypermultiplets in general can become massive and decouple from the low-energy effective action in regions of the moduli space. 

We find that the chiral soft algebra is generated by the operators in Equation \eqref{eq: N2 vector multiplet} which live in a representation of the $\mathcal{N} = 2$ vector multiplet in addition to the following operators which live in a representation of the $\mathcal{N} = 2$ hypermultiplet:
\begin{equation}
    \begin{split}
        \text{$\mathcal{N} = 2$ Hypermultiplet}:& \hspace{15pt}\begin{cases} \hspace{5pt}
        \mathcal{R}^{\lambda,p}_{n,m} \hspace{10pt},\hspace{10pt} \mathcal{R}^{\chi,p}_{n,m} \hspace{10pt},\hspace{10pt}  \mathcal{R}^{q,I,p}_{n,m} \\
        \hspace{5pt} \overline{\mathcal{R}}{}^{\lambda,p}_{n,m} \hspace{10pt},\hspace{10pt} \overline{\mathcal{R}}{}^{\chi,p}_{n,m} \hspace{10pt},\hspace{10pt}  \overline{\mathcal{R}}{}^{q,I,p}_{n,m} \ ,
    \end{cases}
    \end{split}
\end{equation}
for $p = 1,\frac{3}{2},2,...$ and $1-p \leq n \leq p-1$.

The soft algebra with this hypermultiplet included can be partitioned into two pieces. The first piece is exactly Equation \eqref{eq: soft algebra} which describes interactions in the vector multiplet and is completely undeformed in the presence of the new hypermultiplet. In this way, it is an ideal of the full chiral soft algebra. The second set of terms in the chiral soft algebra involve one generator in the vector multiplet and two generators in the hypermultiplet. We list them below:
\begin{equation} 
    \begin{split}
    A\hspace{2pt},\hspace{2pt}q^I\hspace{2pt},\hspace{2pt}\overline{q}^J:& \hspace{16pt} \begin{cases} \hspace{5pt} \Big[\mathcal{R}_{n,m}^{A,p},\overline{\mathcal{R}}{}_{n',m'}^{q,I,p'}\Big] \hspace{2pt} = -\frac{1}{\sqrt{2}}(n_e + \bar \tau n_m)~ \overline{\mathcal{R}}{}_{n+n',m+m'}^{q,I,p+p'-1} \\
    \hspace{5pt} \Big[\mathcal{R}_{n,m}^{A,p},\mathcal{R}_{n',m'}^{q,J,p'}\Big] \hspace{2pt} = \frac{1}{\sqrt{2}}(n_e + \bar \tau n_m)~ \mathcal{R}_{n+n',m+m'}^{q,J,p+p'-1}\\
    \hspace{5pt} \Big[\overline{\mathcal{R}}{}_{n,m}^{q,I,p},\mathcal{R}_{n',m'}^{q,J,p'}\Big] = -\frac{1}{\sqrt{2}}(n_e + \bar \tau n_m)~ \hspace{2pt} \varepsilon^{IJ} \hspace{2pt} \overline{\mathcal{R}}{}_{n+n',m+m'}^{A,p+p'-1}  \end{cases} \\ ~\\
    q^I\hspace{2pt},\hspace{2pt}\lambda\hspace{2pt},\hspace{2pt}\psi^J:& \hspace{16pt} \begin{cases} \hspace{5pt} \Big[\overline{\mathcal{R}}{}_{n,m}^{q,I,p},\mathcal{R}_{n',m'}^{\lambda,p'}\Big] \hspace{2pt} = - \frac{i}{\sqrt{2}}(n_e + \bar \tau n_m)~ \overline{\mathcal{R}}{}_{n+n',m+m'}^{\psi,I,p+p'-1} \\ 
    \hspace{5pt} \Big[\mathcal{R}_{n,m}^{\lambda,p},\mathcal{R}_{n',m'}^{\psi,J,p'}\Big] \hspace{2pt} = \frac{i}{\sqrt{2}}(n_e + \bar \tau n_m)~ \mathcal{R}_{n+n',m+m'}^{q,J,p+p'-1} \\
    \hspace{5pt} \Big[\overline{\mathcal{R}}{}_{n,m}^{q,I,p},\mathcal{R}_{n',m'}^{\psi,J,p'}\Big] = - \frac{i}{\sqrt{2}}(n_e + \bar \tau n_m)~ \hspace{2pt}\varepsilon^{IJ} \hspace{2pt}\overline{\mathcal{R}}{}_{n+n',m+m'}^{\lambda,p+p'-1} \end{cases} \\ ~\\
    \overline{q}^I\hspace{2pt},\hspace{2pt} \chi \hspace{2pt},\hspace{2pt}\psi^J:& \hspace{16pt} \begin{cases} \hspace{5pt} \Big[\mathcal{R}_{n,m}^{q,I,p},\mathcal{R}_{n',m'}^{\chi,p'}\Big] \hspace{1pt} = \frac{i}{\sqrt{2}}(n_e + \bar \tau n_m)~ \overline{\mathcal{R}}{}_{n+n',m+m'}^{\psi,I,p+p'-1} \\
    \hspace{5pt} \Big[\mathcal{R}_{n,m}^{\chi,p},\mathcal{R}_{n',m'}^{\psi,J,p'}\Big] \hspace{2pt} = -\frac{i}{\sqrt{2}}(n_e + \bar \tau n_m)~ \overline{\mathcal{R}}{}_{n+n',m+m'}^{q,J,p+p'-1} \\
    \hspace{5pt} \Big[\mathcal{R}_{n,m}^{q,I,p},\mathcal{R}_{n',m'}^{\psi,J,p'}\Big] = \frac{i}{\sqrt{2}}(n_e + \bar \tau n_m)~ \hspace{2pt} \varepsilon^{IJ} \hspace{2pt} \overline{\mathcal{R}}{}_{n+n',m+m'}^{\chi,p+p'-1} \end{cases} \\   ~\\ 
    \varphi\hspace{2pt},\hspace{2pt}\lambda\hspace{2pt},\hspace{2pt}\chi:& \hspace{16pt} \begin{cases} \hspace{5pt} \Big[\mathcal{R}_{n,m}^{\varphi,p},\mathcal{R}_{n',m'}^{\lambda,p'}\Big] =  \frac{i}{\sqrt{2}}(n_e + \bar \tau n_m)~ \overline{\mathcal{R}}{}_{n+n',m+m'}^{\chi,p+p'-1} \\
    \hspace{5pt} \Big[\mathcal{R}_{n,m}^{\lambda,p},\mathcal{R}_{n',m'}^{\chi,p'}\Big] = -\frac{i}{\sqrt{2}}(n_e + \bar \tau n_m)~ \overline{\mathcal{R}}{}_{n+n',m+m'}^{\varphi,p+p'-1} \\
    \hspace{5pt} \Big[\mathcal{R}_{n,m}^{\varphi,p},\mathcal{R}_{n',m'}^{\chi,p'}\Big] = \frac{i}{\sqrt{2}}(n_e + \bar \tau n_m)~ \overline{\mathcal{R}}{}_{n+n',m+m'}^{\lambda,p+p'-1} \end{cases} \\~ \\
    A\hspace{2pt},\hspace{2pt}\lambda\hspace{2pt},\hspace{2pt}\bar \lambda:& \hspace{16pt} \begin{cases} \hspace{5pt} \Big[\mathcal{R}_{n,m}^{A,p},\mathcal{R}_{n',m'}^{\lambda,p'}\Big] = \frac{1}{\sqrt{2}}(n_e + \bar \tau n_m)~ \mathcal{R}_{n+n',m+m'}^{\lambda,p+p'-1}\\
    \hspace{5pt} \Big[\mathcal{R}_{n,m}^{A,p},\overline{\mathcal{R}}{}_{n',m'}^{\lambda,p'}\Big] = \frac{1}{\sqrt{2}}(n_e + \bar \tau n_m)~ \overline{\mathcal{R}}{}_{n+n',m+m'}^{\lambda,p+p'-1} \\ \hspace{5pt} \Big[\mathcal{R}_{n,m}^{\lambda,p},\overline{\mathcal{R}}{}_{n',m'}^{\lambda,p'}\Big]  = - \frac{1}{\sqrt{2}}(n_e + \bar \tau n_m)~ \overline{\mathcal{R}}{}_{n+n',m+m'}^{A,p+p'-1}
    \end{cases} \\ ~\\
    A\hspace{2pt},\hspace{2pt}\chi\hspace{2pt},\hspace{2pt}\bar \chi:& \hspace{16pt} \begin{cases} \hspace{5pt}  \Big[\mathcal{R}_{n,m}^{A,p},\mathcal{R}_{n',m'}^{\chi,p'}\Big] = \frac{1}{\sqrt{2}}(n_e + \bar \tau n_m)~ \mathcal{R}_{n+n',m+m'}^{\chi,p+p'-1}\\ \hspace{5pt}\Big[\mathcal{R}_{n,m}^{A,p},\overline{\mathcal{R}}{}_{n',m'}^{\chi,p'}\Big] = \frac{1}{\sqrt{2}}(n_e + \bar \tau n_m)~ \overline{\mathcal{R}}{}_{n+n',m+m'}^{\chi,p+p'-1} \\
    \hspace{5pt} \Big[\mathcal{R}_{n,m}^{\chi,p},\overline{\mathcal{R}}{}_{n',m'}^{\chi,p'}\Big]  = - \frac{1}{\sqrt{2}}(n_e + \bar \tau n_m)~ \overline{\mathcal{R}}{}_{n+n',m+m'}^{A,p+p'-1}\hspace{2pt}. \\
    \end{cases}\\    
\end{split}
\label{eqn: hypermultiplet soft algebra}
\end{equation}
It is readily verified that this chiral soft algebra is $SL_2(\mathbb{Z})$ covariant and satisfies the Jacobi identity, as do all such algebras with $\mathcal{N} \geq 1$ bulk supersymmetry \cite{Ball:2023qim}.

\section{The Goldstone Sector}
\label{sec:Goldstones}
The  charge lattice of BPS particles varies over the moduli space ${\mathcal{M}}_\phi$ and has played a central role in the study of its properties \cite{seiberg1994electric,tachikawa2013n, labastida2005topological, Bilal:1995hc}. The charges are sourced by Wilson-`t Hooft lines, which have a celestial realization as exponentials of 2D Goldstone modes  \cite{nande2018soft}. 
The $\Delta =1$ soft symmetry generators are symplectically paired,  either via the 4D Klein-Gordon or 2D BPZ inner products, with these Goldstone modes. Together, this symplectic pair generates the (photon sector of the) soft $S$-matrix. In this section we will show that they can be described by a pair of free 2D bosons  associated to a signature  $(2,2)$ Narain lattice with modulus  $\tau$.

For brevity, in this section we do not consider either $\Delta<1$ soft modes\footnote{The Goldstone partners of other soft symmetry generators with $k<1$ are discussed for example in  \cite{Freidel:2022skz, Crawley:2023brz, Freidel:2023gue}. Other constructions of Goldstone operators can be found in \cite{Donnay:2018neh, Himwich:2020rro,Arkani-Hamed:2020gyp, Kapec:2021eug,Nguyen:2021ydb, donnay2022goldilocks}.} or  the superpartners of the $\Delta=1$ soft photon modes. This would extend the operator spectrum below to supermultiplets and allow for nonzero spin. 

\subsection{Minkowski Space}
The electromagnetic charges of particles lie on a charge lattice $\Gamma$ defined by
\begin{equation}
    \Gamma = \Big\{e(n_e + \tau n_m) \hspace{2pt}\Big|\hspace{2pt} n_e,n_m \in \mathbb{Z}\Big\} \subset \mathbb{C}\hspace{2pt},
    \label{eqn: charge lattice}
\end{equation}
where the real and imaginary parts of  vectors $\gamma \in \Gamma$ are  electric and magnetic charges respectively. We denote by
$ \mathcal{W}[\gamma](z,\bar z)$ the 
 Wilson ('t Hooft) line operators introduced in   \cite{nande2018soft} which source these charges and generate the conformally soft part of the S-matrix.  The leading soft photon theorem implies the $SL_2(\mathbb{Z})$ covariant 2D OPEs 
\begin{equation}
    \begin{split}
        2\pi
 J(z) \hspace{2pt} \mathcal{W}[\gamma](w, \bar w) &\sim -\frac{{\bar\gamma}}{z-w} \hspace{2pt} \mathcal{W}[\gamma](w, \bar w ) \ , \hspace{30pt} 2\pi \bar J(\bar z) \hspace{2pt} \mathcal{W}[\gamma](w, \bar w ) \sim  \frac{{\gamma}}{\bar z-\bar w } \hspace{2pt} \mathcal{W}[\gamma](w, \bar w ) . \\ 
    \end{split}\label{opa}
\end{equation}
Here  $J(z)= \tfrac{1}{2\pi}\sum_m \mathcal{R}^{A,1}_{0,m} \hspace{1pt} z^{-m-1}$ denotes the chiral soft current and $\bar J(\bar z)$  a corresponding   antichiral soft current.\footnote{Note we do not employ a chiral expansion in this section in which the $1/\bar z$ poles appearing in Equation \eqref{opa} would be absent.} 
In this paper we work at tree level (of the low-energy effective action) in which Wilson lines do not interact and the $\mathcal{W}\mathcal{W}$ OPE is 
\begin{equation}
    \mathcal{W}[\gamma](z,\bar z) \hspace{2pt} \mathcal{W}[\gamma'](w, \bar w) = \hspace{3pt}:\mathcal{W}[\gamma](z,\bar z) \mathcal{W}[\gamma'](w, \bar w ):.\label{opb}
\end{equation}
This implies that the tree-level spin and conformal dimension of $\mathcal{W}$ both vanish. Equation \eqref{opb} is  corrected at one loop where the cusp anomalous dimension imparts  a conformal dimension for  $\mathcal{W}$. Such loop effects were incorporated in  \cite{nande2018soft,Arkani-Hamed:2020gyp}, it would be of interest to so expand the discussion here. 

The OPEs \eqref{opa} and \eqref{opb}  are defining equations for the (tree-level) soft S-matrix. They may be given a free field realization \cite{nande2018soft}.\footnote{The realization here is in the spirit of \cite{nande2018soft} but differs in detail.} Let us define two chiral and two antichiral bosons  $\big({\Phi}(z),  \bar{\Phi}(\bar z), \Psi(z), \bar{\Psi}(\bar z) \big)$ with off-diagonal OPEs
\begin{equation}
    \begin{split}
        \Psi(z) {\Phi}(w) &\sim - {\ln(z-w)\over 4 \pi^2}\hspace{86pt}  \bar{\Psi}(\bar z) \bar{\Phi}(\bar w) \sim {\ln(\bar z-\bar w) \over 4 \pi^2} \\
        {\Phi}(z) {\Phi}(w) &\sim 0 \hspace{137pt} \bar{\Phi}(\bar z) \bar{\Phi}(\bar w) \sim 0 \\
        \Psi(z) \Psi(w) &\sim 0 \hspace{136pt} \bar{\Psi}(\bar z) \bar{\Psi}(\bar w) \sim 0.
        \label{eqn: Free boson OPE}
    \end{split}
\end{equation}
From these we may construct the Wilson lines
\begin{equation}
    \mathcal{W}[\gamma] = \hspace{2pt}:\exp(-2\pi i \bar\gamma \hspace{2pt} {\Phi}- 2\pi i  \gamma \hspace{2pt} \bar{\Phi}): \hspace{2pt},
    \label{eqn: Wilson line}
\end{equation}
and soft currents
\be J(z)=i\p \Psi(z)\ , ~~~~~\bar J(\bar z)= 
i\bar \p \bar \Psi(\bar z).\label{dxz}\ee Equation \eqref{eqn: Free boson OPE} then reproduces the defining relations  \eqref{opa} and \eqref{opb}.

From \eqref{eqn: Wilson line} we see that $(\Phi, \bar \Phi)$ 
are periodically identified under shifts  in the dual lattice $\Gamma^\vee$\begin{equation}
 {\Phi}(z) \sim {\Phi}(z)+ {\gamma}^\vee , \hspace{25pt} \bar{\Phi}(\bar z) \sim \bar{\Phi}(\bar z) +\bar\gamma^\vee, \hspace{35pt} \gamma^\vee \in \Gamma^\vee   ,\label{eqn:periodicity}\end{equation}
where the  lattice $\Gamma^\vee$ dual to $\Gamma$ consists of those vectors $\gamma^\vee \in \mathbb{C}$ with
\begin{equation}
    (\gamma,\gamma^\vee) \equiv \gamma \bar{\gamma}^\vee + \bar \gamma \gamma^\vee \in \mathbb{Z} \hspace{30pt} \text{for all} ~ ~ \gamma \in \Gamma\hspace{2pt}.
    \label{eqn: quantization condition}
\end{equation}
 One finds that 
\begin{equation}
    \Gamma^\vee =  \Big\{\frac{e}{8\pi i} \big(\q + \tau \s) \hspace{2pt}\Big|\hspace{2pt} \q,\s \in \mathbb{Z}\Big\}  \subset \mathbb{C}\hspace{2pt}.
    \label{eqn: dual lattice}
\end{equation}

$(\Phi,\bar \Phi)$ are the Goldstone bosons of large gauge  symmetry, under which they transform as 
\be\label{lgat} \Phi(z) \to \Phi(z)+\epsilon(z),~~~~\bar \Phi(z) \to \bar \Phi(\bar z)+\bar \epsilon(\bar z),\ee 
where $\bar \epsilon(\bar z)=(\epsilon(z))^*$ and the real (imaginary) part of  $\epsilon$ generates  large electric (magnetic) gauge transformations \cite{Strominger:2015bla}. The identification \eqref{eqn:periodicity} implies that transformations with $\epsilon \in \Gamma^\vee$ are trivial. 
Large gauge transformations are generated by contour integrals of the  one form $Jdz+\bar Jd\bar z$.  For  $\epsilon(z)$ holomorphic  inside a contour $C$ around $w$, \eqref{eqn: Free boson OPE} gives 
\begin{equation}
    \begin{split}
 2\pi \oint_C\epsilon  J~  \Phi(w) \sim {\epsilon (w)},\\~~~~ 2\pi \oint_C\bar \epsilon  \bar J~\bar  \Phi(\bar w) \sim {\bar \epsilon (\bar  w) }.  \end{split}
\end{equation}

It is natural to also consider exponentials of the soft bosons $(\Psi,\bar \Psi)$ as well as the Goldstone bosons. We denote these
\be \mathcal{U}[\beta]=:\exp(-2\pi i \beta \hspace{2pt} \Psi-2\pi i \bar\beta \hspace{2pt} \bar{\Psi}): .\ee 
Their OPE with the Wilson line operators is 
\be \mathcal{U}[\beta](z,\bar z)  \mathcal{W}[\gamma](w,\bar w)=(z-w)^{\beta \bar\gamma}(\bar z-\bar w)^{-\bar\beta \gamma}:\mathcal{U}[\beta](z,\bar z)  \mathcal{W}[\gamma](w,\bar w):.\label{sbn}\ee
Comparison with \eqref{lgat}  identifies them as generators of finite large gauge transformations
\be  2\pi \epsilon_z( w)= i \beta \ln (z-w),~~ 2\pi \bar \epsilon_{\bar z} (\bar w)=-i\bar \beta\ln(\bar z-\bar w).\ee
Single valuedness of \eqref{sbn}, or equivalently 
locality on the celestial sphere, then requires 
\be  \beta \bar \gamma+\bar \beta \gamma\in {\mathbb Z}. \ee
This implies that the branch cut discontinuity  
\be \Delta \epsilon_z =-\beta \in  \Gamma^\vee, \ee is a trivial gauge transformation, as well as the periodic identification 
\be \Psi(z) \sim \Psi(z) + \bar\gamma , \hspace{25pt}  \bar{\Psi}(\bar z) \sim  \bar{\Psi}(\bar z) + \gamma,\hspace{31pt} \gamma \in \Gamma. \ee

Of course, we can consider more general  operators by exponentiating the soft boson as well as the Goldstone boson. A more general set of mutually local  operators are  
\begin{equation}
    \mathcal{V}[\gamma,\beta] = \hspace{2pt}:\exp\big(-2\pi i \bar\gamma \hspace{1pt} {\Phi} - 2 \pi i \gamma \hspace{1pt} \bar{\Phi} -2\pi i \beta \hspace{1pt} \Psi - 2\pi i \bar\beta\hspace{1pt} \bar{\Psi} \big):.
\end{equation}
The OPE between such  operators is \be \mathcal{V}[\gamma,\beta](z,\bar z)\mathcal{V}[\gamma',\beta'](w,\bar w)=(z-w)^{\bar\gamma \beta^\prime +\bar\gamma^\prime \beta} (\bar z-\bar w)^{-\gamma \bar\beta^{\prime} - {\gamma}^\prime \bar \beta}: \mathcal{V}[\gamma,\beta](z,\bar z)\mathcal{V}[\gamma',\beta'](w,\bar w): \hspace{2pt} .
\ee
We see that these operators are mutually local provided that 
\be \label{spn} (\gamma,\beta; \gamma',\beta')\equiv \gamma\bar\beta'+\beta\bar \gamma'+\bar\gamma \beta'+\bar\beta \gamma' \in {\mathbb{Z}} \ ,\ee
which is implied by the fact that $\beta$ and $\gamma$ lie in dual lattices. In  particular at leading order for $z\to w$, we have\footnote{The algebra of the $\mathcal{V}$'s is like a vertex operator algebra except that it is not unitary.}
\be \mathcal{V}[\gamma,\beta](z,\bar z)\mathcal{V}[-\gamma,-\beta](w,\bar w)=
(z-w)^{-2\bar\gamma \beta } (\bar z-\bar w)^{2\gamma \bar\beta} + \ldots
\ee
It follows that the conformal dimension and spin of $\mathcal{V}[\gamma, \beta]$ are :
\begin{equation}
    \begin{split}
        \Delta\Big(\mathcal{V}[\gamma,\beta]\Big) &=  \bar\gamma \beta- \gamma \bar\beta\\
        &= \frac{2}{\tau - \bar\tau} \Big(n_e \q + \frac{1}{2}\left(\tau + \bar\tau\right)(n_e \s + n_m \q) + \tau\bar\tau n_m \s \Big) \hspace{2pt},\\
        s\Big(\mathcal{V}[\gamma,\beta]\Big) &= \gamma \bar\beta +\bar\gamma \beta\\
        &= n_e \s - n_m \q \in \mathbb{Z}.
        \label{eqn: dimension and spin}
    \end{split}
\end{equation}
  The  inner product  $(\gamma,\beta; \gamma'\beta')$ on $\Gamma \oplus \Gamma^\vee$ appearing in \eqref{spn} defines an even,  self-dual, Lorentzian signature $(2,2)$ Narain lattice. The 2D bosons $(\Phi,\bar \Phi,\Psi, \bar \Psi)$ are coordinates on the Narain torus defined by this Narain lattice.
This resembles  the Narain construction for two string worldsheet bosons compactified on a 2-torus. The expression \eqref{eqn: dimension and spin} for the conformal weight $\Delta$ however is purely imaginary\footnote{This corresponds to power law rather than oscillatory behavior as null infinity is approached. Below we will see weights become real in Klein space.} and not the one appearing in the standard worldsheet construction.  

   The operators $\mathcal{V}[\gamma,\beta]$ and their descendants generate the electromagnetic part of the 
   soft S-matrix. 

\subsection{Comments}
In general there appears to be a rich structure of operator variation on the moduli space. While we will not pursue this in detail here, it is interesting to ask whether there are points in the moduli space for which any of the $\mathcal{V}[\gamma,\beta]$ can appear in the chiral algebra discussed above.  The simplest such operators have $2 \bar h =\Delta-s=0$. Since $\Delta$ is purely  imaginary (corresponding to a real Rindler boost energy) this is possible only for $\Delta=0$. Such operators are plentiful, for example at any point where $\tau + \bar\tau={\theta \over \pi}=0$. A more general possibility is $\Delta=0$, $h=n={{-}}\bar h$ with $n$ a positive half-integer. In this case the operator transforms in the $(2n+1,-2n-1)$ representation of $SL_2(\mathbb{C})$, which can be realized as a chiral field with $2n+1$ components \cite{Guevara:2021abz}. The subleading soft photons and gravitons are examples of operators  in this representation with $n={1\over 2}$ and $n=1$. The next subsection briefly discusses Klein space where the conformal dimensions are all real providing  a natural place to discuss chiral algebras. 

The appearance of massless states at various points  is a key ingredient in the analysis of the scalar moduli space $\mathcal{M}_{\phi}$. Similarly, one may expect that the appearance of extra chiral operators at various points on $\mathcal{M}_\phi$ may provide an organizing principle for understanding the geometry of the full moduli space $\mathcal{M}$. 

We wish to stress that, although they do have a local action on the Hilbert space, we do not expect all of these operators to create physical scattering states.  Wilson line operators in general provide only the charged  dressings of matter fields.  In most examples, stable physical scattering states occupy only a small portion of the charge lattice. For example, in a neighborhood of the massless monopole singularity 
of \cite{seiberg1994electric}, one can pick a duality frame where only charges $\pm(1,0)$ and $\pm(1,1)$ are realized by stable particles.   We also note that there are a large number of dimension zero operators: for example anything with $\gamma=0$ or $\beta=0$.   A complete picture of the physical spectrum will in addition require consideration of IR divergences and the cusp anomalous dimension which shifts $\Delta$ \cite{Kapec:2017tkm,nande2018soft,Arkani-Hamed:2020gyp}. We leave this to future work!

\subsection{Klein Space}
\label{sec: Vertex Ops Klein}
Upon Wick rotating to Klein space, $z$ and $\bar z$ become real and independent variables, and $\tau$ and $\bar \tau$ become real, independent parameters which are related to the electric coupling and $\theta$-angle in Minkowski space via $\tau + \bar \tau = \theta/\pi$ and $\tau - \bar \tau = 8\pi/e^2$. Discussion of nonperturbative aspects of split-signature $\mathcal{N} = 2$ moduli spaces can be found in \cite{Dijkgraaf:2016lym}.

The analytically continued charge  lattice and its dual  become
\begin{equation}
    \begin{split}
        \Gamma &= \Big\{\hspace{6pt} \sqrt{\frac{8\pi}{\tau - \bar \tau}}\big( n_e + \tau n_m,n_e + \bar \tau n_m\big)\Big|n_e,n_m \in \mathbb{Z}\Big\}\subset \mathbb{R}^2\hspace{2pt} , \\
        \Gamma^\vee &= \Big\{\frac{1}{8\pi}\sqrt{\frac{8\pi}{\tau - \bar \tau}}\big(\q + \tau \s, -\q - \bar \tau \s \big)\Big|\hspace{2pt} \q,\s \in \mathbb{Z}\Big\}\hspace{2pt}\subset \mathbb{R}^2\hspace{2pt} ,
    \end{split}
\end{equation}
where   $\Gamma^\vee$ is defined via $\gamma \bar{\gamma}^\vee +  \bar{\gamma} \gamma^\vee \in \mathbb{Z}$ as before, except that $\gamma$ and $\bar \gamma$ are now real independent variables.

The free field construction for the conformally soft sector is as in the previous section with $(z, \bar z,\gamma, \bar \gamma, \Phi, \bar \Phi, \Psi, \bar \Psi, \tau, \bar \tau)$ all real and independent. The dimension and spin of the analytically continued vertex operator $\mathcal{V}[\gamma,\beta]$ is
\begin{equation}
    \begin{split}
        \Delta\Big(\mathcal{V}[\gamma,\beta]\Big) &=  \bar\gamma \beta - \gamma \bar\beta \\
        &= \frac{2}{\tau - \bar\tau} \Big(n_e \q + \frac{1}{2}\left(\tau + \bar\tau\right)(n_e \s + n_m \q) + \tau\bar\tau n_m \s \Big) \hspace{2pt},\\
        s\Big(\mathcal{V}[\gamma,\beta]\Big) &= \bar\gamma \beta + \gamma \bar\beta\\
        &= n_e \s - n_m \q \in \mathbb{Z}.
        \label{eqn: klein dimension and spin}
    \end{split}
\end{equation}
Note that in Klein space $\Delta$  is real.  

At certain points in the Kleinian moduli space, one can find points for $n_e,n_m \in \mathbb{Z}$ such that $\gamma = 0$ while $\bar \gamma \neq 0$. This happens, for example, whenever $\tau \in \mathbb{Q}$. 
At such a point, the corresponding vertex operators have dimension and spin
\begin{equation}
    \Delta = s = n_e \s - n_m \q \in \mathbb{Z}.
\end{equation}
These potentially provide interesting new generators of the chiral soft algebra. Chiral soft algebras are naturally considered in Klein space because, as mentioned in the introduction, the holomorphic expansion treats  $z$ and $\bar z$  as independent which is essentially the same as analytic continuation from Minkowski to Klein space. 

\section{Moduli Space Singularities} 
\label{sec:strong coupling}
The analysis of singularities at certain points in the moduli space $\mathcal{M}_{\phi}$ has been a key ingredient in the study of 
 $\mathcal{N} = 2$ gauge theories. Often, these are characterized by the appearance of extra massless hypermultiplets which resolve the singularity.   In this section, we describe the chiral soft algebras, including  the dimensions of $\mathcal{V}[\gamma,\beta]$, at and near the singularities and find that new generators appear.

 We illustrate the questions  here with the canonical Seiberg-Witten example of the $SU(2)$ $\mathcal{N} = 2$ gauge theory.   This has two strong coupling singularities at which a monopole ($u=1$) or dyon ($u=-1)$ become massless and a third weak-coupling singularity at $u=\infty$. The coupling constant behaves as 
 \begin{equation}
    \mu \xrightarrow{\hspace{5pt}\text{monopole}\hspace{5pt}} \infty \ , \hspace{40pt} \mu \xrightarrow{\hspace{14pt}\text{dyon}\hspace{14pt}} \infty \ , \hspace{40pt} \mu \xrightarrow{\text{weak coupling}} 0.
\label{sng}\end{equation}
Encircling a singularity in the moduli space induces a monodromy on the special coordinates (see Figure \ref{fig: bundle})
\begin{equation}
    \binom{a_D}{a} \longmapsto \binom{a_D'}{a'} = M \binom{a_D}{a} , 
    \label{eqn: monodromy special coordinates}
\end{equation}
where $M \in  SL_2(\mathbb{Z})$.  At $u = 1$, $u = -1$, and $u = \infty$ the monodromies are 
\begin{equation}
M_{1}=\binom{
\hspace{4pt} 1 \hspace{13pt} 0}{-2 \hspace{10pt} 1}
\ ,\hspace{30pt} M_{-1}=\binom{
\hspace{1pt}3 \hspace{13pt} 2}{-2 \hspace{5pt} -1}\ ,\hspace{30pt} M_{\infty}=\binom{\hspace{-7pt}
-1 \hspace{10pt} 2}{\hspace{4pt} 0 \hspace{5pt} -1} \ .
\end{equation} 

Let us now consider the behavior of the CCFT and chiral algebra at these points. Since $\mu$ is proportional to the structure constants of the algebra, from \eqref{sng} we see that at the weak coupling singularity the chiral soft algebra becomes abelian. This makes sense because in the bulk, the weak-coupling singularity coincides with the limit where the vector multiplet becomes free.

On the other hand, at $u=\pm 1$ these structure constants diverge, and the chiral soft  algebra is singular. Of course, using the 4D bulk description,  we can go to the resolved picture near the singularity by including the emerging massless particles. This brings along with it new chiral generators and there is a corresponding smooth, resolved chiral soft algebra taking the same general form as Equations \eqref{eq: soft algebra} and \eqref{eqn: hypermultiplet soft algebra} but with modified, non-singular values of $\mu$ and $\tau$. However we do not have an understanding of how the new generators resolve the emerging degeneracies of the 2D chiral soft algebra without reference to the 4D bulk dynamics. In principle, there may be  a fully 2D description in which 
the divergent algebra near the singularity can be derived from the resolved algebra. This may require incorporating quantum effects which shift the conformal dimensions at one loop.

It  is interesting to study the behavior of the operators $\mathcal{V}[\gamma,\beta]$ on $\mathcal{M}_{\phi}$.  
These operators  have quantized spin which is therefore constant, but the dimension degenerates at the strong coupling singularities.
As $u\to 1$ ($u\to -1$), one has  $\tau\to 0$ ($\tau\to -1$).  In either case  $\tau-\bar\tau\to 0$, so the charge lattice $\Gamma$ is degenerating, with lattice points on the imaginary axis approaching one another. Labelling the vertex operators by their quantum numbers $(n_m,n_e,n_m^\vee,n_e^\vee)$, the conformal dimensions behave as \begin{equation}
    \begin{split}
        \Delta\Big(\mathcal{V}[n_m,n_e,\s,\q]\Big) &~ \xrightarrow{\hspace{7pt}\text{monopole}\hspace{7pt}} ~ \frac{2}{\tau - \bar\tau} n_e \q \hspace{2pt}, \\
        \Delta\Big(\mathcal{V}[n_m,n_e,\s,\q]\Big) &~ \xrightarrow{\hspace{16pt}\text{dyon}\hspace{16pt}} ~ \frac{2}{\tau - \bar\tau}(n_e-n_m)(\q - \s)\hspace{2pt}.
        \label{eqn: dimension and spin_limit}
    \end{split}
\end{equation}On the other hand, as we approach the weak coupling singularity
\begin{equation}
    \Delta\Big(\mathcal{V}[n_m,n_e,\s,\q]\Big) ~ \xrightarrow{\text{weak coupling}} ~ \frac{2\tau\bar\tau}{\tau - \bar\tau} n_m \s  \hspace{2pt}, \hspace{67pt}
    \label{eqn: dimension and spin (weak coupling)}
\end{equation}
where the ratio $\frac{\tau\bar\tau}{\tau - \bar\tau} \hspace{1pt}  \rightarrow \infty$ as $u\to\infty$. In any of these cases, dimensions of generic operators all go either to zero or infinity. 

The $\mathcal{V}[n_m,n_e,n_m^\vee,n_e^\vee]$ operators have monodromy around the singular points. One can equivilently view the monodromy group as acting on the quantum numbers $(n_m,n_e,n_m^\vee,n_e^\vee)$ as
\begin{equation}
    \begin{split}
        (n_m',n_e') = (n_m,n_e) \hspace{1pt} M \ , \hspace{50pt}
      (n_m'^{\hspace{2pt}\vee},n_e'^{\hspace{2pt}\vee}) = (\s,\q)\hspace{1pt} M \ ,
    \end{split}
    \label{eqn: monodromy quantum numbers}
\end{equation}
where under the monodromy
\begin{equation} \label{eq:vertex_monodromy}
    \mathcal{V}[n_m,n_e,\s,\q] ~ \longmapsto ~ \mathcal{V}[n_m',n_e',n_m'^{\hspace{2pt}\vee},n_e'^{\hspace{2pt}\vee}] \ .
\end{equation}
Quantum numbers $(n_m,n_e)$ that are monodromy eigenvectors  with unit eigenvalue are of special interest. For example, the quantum numbers for a monopole, $(1,0)$, is a left-eigenvector of $M_{1}$ with this property, which is why a monopole (rather than some other BPS particle) condenses at $u = 1.$ One can similarly verify that the dyon quantum numbers, $(1,1)$, are a left-eigenvector of $M_{-1}$ with unit eigenvalue. One can also consider pairs $(\s,\q)$ which are left-eigenvectors of $M_{\pm 1}$ with unit eigenvalue. We summarize these results below
\begin{equation}
    \begin{split}
        M_{1}&: \hspace{5pt} (n_m,n_e) = (1,0) \ , \hspace{30pt} (\s,\q) = (1,0)\ ,\\
        M_{-1}&: \hspace{5pt} (n_m,n_e) = (1,1)\ , \hspace{30pt} (\s,\q) = (1,1) \ .\\
    \end{split}
\end{equation}
By contrast, $M_{\infty}$ has no left-eigenvector with eigenvalue $1$. This is why the weak-coupling singularity cannot simply be interpreted as a charged particle becoming massless. On the other hand, $M_{\infty}$ does have left-eigenvectors with eigenvalue $-1$:
\begin{equation}
    M_{\infty}: \hspace{5pt} (n_m,n_e) = (0,1)\ , \hspace{30pt} (\s,\q) = (0,1) \ .\\
\end{equation}
In all instances, each pair of eigenvectors encode a sublattice of charge vectors $(n_m,n_e,\s,\q)$.
\begin{equation}
    \begin{split}
        \Xi_{1} &= \mathbb{Z}\hspace{2pt} (1,0) \oplus \mathbb{Z}(1,0)\ ,\\
        \Xi_{-1} &= \mathbb{Z}\hspace{2pt} (1,1) \oplus \mathbb{Z}(1,1)\ , \\
        \Xi_\infty &= \mathbb{Z}\hspace{2pt}(0,1) \oplus \mathbb{Z} \hspace{2pt}(0,1) \ .
    \end{split}
\end{equation}
At the $u = \pm 1$ singularities, only the  operators $\mathcal{V}[n_m,n_e,\s,\q]$ with coefficients $(n_m,n_e,\s,\q) \in \Xi_{\pm 1}$ will be invariant under the monodromy $M_{\pm 1}$ and well-defined at the singularity. One can verify from \eqref{eqn: dimension and spin} that all such operators have $\Delta = 0$, $s = 0$ at the singularity. On the other hand,  operators which do not have coefficients in this sublattice generically have $\Delta \rightarrow \infty.$

At the $u = \infty$ singularity, operators with coefficients in $\Xi_\infty$ will not be invariant under $M_{\infty}$. Rather, $\mathcal{V}[n_m,n_e,\s,\q] \mapsto$ $\mathcal{V}[-n_m,-n_e,-\s,-\q]$. On the other hand, there is a $\mathbb{Z}_2$ subgroup of the global $R$-symmetry group which acts on quantum numbers in precisely this way \cite{seiberg1994electric}. Linear combinations of operators which transform trivially under this $R$-symmetry group will be invariant under $M_{\infty}$ and have $\Delta = 0, s = 0$ at the weak coupling singularity.

Though the lattices $\Xi_{1},\Xi_{-1},$ and $\Xi_{\infty}$ are all composed of infinitely many vectors, an arbitrary element can be  generated by exactly 4 vertex operators. For example, at the monopole point, all such elements take  the form $\mathcal{V}[p,0,q,0]$. Two such operators have the OPE $\mathcal{V}[p,0,q,0] \mathcal{V}[p',0,q',0] = \mathcal{V}[p+p',0,q+q',0].$ Thus, the most general invariant operator can be obtained by taking successive OPEs of the following four generators
\begin{equation}
    \mathcal{V}[1,0,0,0] \ ,\hspace{10pt} \mathcal{V}[-1,0,0,0] \ ,\hspace{10pt} \mathcal{V}[0,0,1,0] \ ,\hspace{10pt} \mathcal{V}[0,0,-1,0]\hspace{2pt}.
\end{equation}

In this way, there are only 4 operators which are fundamental objects in the theory at the strong coupling singularities (2 at the weak coupling singularity). Such operators always have spin $s = 0$, but their dimension is non-zero away from the strong coupling singularities.

Ultimately a purely 2D description of the singularity and its resolution, which does not directly refer to the 4D  picture, would be desirable. We hope to have taken a step in that direction. 

\section*{Acknowledgements}
The authors are grateful for helpful conversations with Dan Kapec, Walker Melton, Sruthi Narayanan and Atul Sharma. This work was supported by DOE Grant de-sc/0007870,  the Simons Collaboration for Celestial Holography,  an NSERC  PGSD fellowship and an Ashford Fellowship to EC and NSF GRFP DGE1745303 to AT.

\appendix

\section{Conventions for Mode Expansions and Primary Operators}
\label{appendix: conventions}

We consider the following free-field mode-expansions for the canonically normalized photon field, $A^\mu$, two-component Weyl spinor field, $\psi_I$, and complex scalar field $\varphi$ in the $\mathcal{N} = 2$ vector multiplet
\begin{equation}
    \begin{split}
        A_\mu(x) &= \frac{1}{e} \sum_{\pm} \int \frac{d^3p}{(2\pi)^3} \frac{1}{2p^0} \Big((\varepsilon^\pm_\mu(p))^*\hspace{2pt} a^{A}_{\pm}(p) e^{ip \cdot x} + \varepsilon^\pm_\mu(p) \hspace{2pt}\big(a^A_{\pm}(p)\big)^\dagger e^{-ip\cdot x}\Big), \\
        \psi^{I}_{\alpha}(x) &= \frac{1}{e} \int \frac{d^3 p}{(2\pi)^3} \frac{| p\rangle_{\alpha}}{2p^0}\Big(a^{\bar\psi, I}_{+}(p) e^{ip \cdot x} + \big(a^{\psi, I}_{-}(p)\big)^\dagger e^{-ip \cdot x}\Big), \\
        \varphi(x) &= \frac{1}{e} \int\frac{d^3 p}{(2\pi)^3} \frac{1}{2p^0} \Big(a^{\overline{\varphi}}(p) e^{i p \cdot x} + \big(a^\varphi(p)\big)^\dagger e^{-ip \cdot x}\Big).
    \end{split}
\end{equation}
where the $\pm$ subscript on the photon and fermion mode operators label the corresponding particle's helicity. One can read off the mode expansion for $\overline{\varphi}$ and $\bar{\psi}^I$ by noting that these fields are related to $\varphi$ and $\psi_I$ by Hermitian conjugation.  As in \cite{guevara2021holographic} we consider in this paper only outgoing soft particles, One defines the (outgoing) conformal primary operators as
\begin{equation}
    \begin{split}
        \mathcal{O}^{A}_{\Delta}(z,\bar z) &\equiv \int_0^\infty \frac{d\omega}{\omega} \omega^\Delta \hspace{2pt} a^A_{+}(\omega,z,\bar z) ,  \hspace{44pt} \overline{\mathcal{O}}^{A}_{\Delta}(z,\bar z) \equiv \int_0^\infty \frac{d\omega}{\omega} \omega^\Delta \hspace{2pt} a^A_{-}(\omega,z,\bar z), \\
        \mathcal{O}^{\psi,I}_{\Delta}(z,\bar z) &\equiv \int_0^\infty \frac{d\omega}{\omega} \omega^\Delta \hspace{2pt} a^{\bar \psi,I}_{+}(\omega,z,\bar z) ,\hspace{33pt} \overline{\mathcal{O}}^{\psi,I}_\Delta(z,\bar z) \equiv \int_0^\infty \frac{d\omega}{\omega} \omega^\Delta \hspace{2pt} a^{\psi,I}_{-}(\omega,z,\bar z),\\
        \mathcal{O}^\varphi_\Delta(z,\bar z) &\equiv \int_0^\infty \frac{d\omega}{\omega} \omega^\Delta \hspace{2pt} a^{\overline{\varphi}}(\omega,z,\bar z) , \hspace{46pt}\overline{\mathcal{O}}^\varphi_\Delta(z,\bar z) \equiv \int_0^\infty \frac{d\omega}{\omega} \omega^\Delta \hspace{2pt} a^{\varphi}(\omega,z,\bar z).
    \label{eqn: Mellin appendix}
    \end{split}
\end{equation}

Operators in the unbarred supermultiplet are Mellin transforms of positive helicity annihilation operators (other than the scalars which have helicity 0), and therefore have $s \geq 0.$ Likewise, operators in the barred supermultiplet have $s \leq 0.$ Finally, note that the R-symmetry index $I$ on the operator $\mathcal{O}^{\psi,I}_{\Delta}$ can be raised and lowered with $\varepsilon^{IJ}$, though we will always work in a convention where it is raised.  We will also exclusively work with the Mellin transform of annihilation operators (these give the so-called ``out'' operators in the CCFT).

In the monopole hypermultiplet, the free field mode expansion for the Weyl spinor fields $\lambda$ and $\chi$ mirrors that of $\psi^I$ while the free field mode expansion for the complex scalar field doublet $q^I$ mirrors that of $\varphi$. Conformal primary operators in the monopole hypermultiplet are obtained via Mellin transforms mirroring \eqref{eqn: Mellin appendix}.

For Pauli matrices, we always use the convention: $\sigma^{\mu}_{\alpha \dot{\alpha}} = (\mathbf{1}_{\alpha \dot{\alpha}},\sigma^i_{\alpha \dot{\alpha}})$ and $\overline{\sigma}^{\mu \dot{\alpha} \alpha} = \varepsilon^{\dot{\alpha} \dot{\beta}} \varepsilon^{\alpha \beta} \sigma^{\mu}_{\beta \dot{\beta}} = (\mathbf{1}^{\dot{\alpha} \alpha},-\sigma^{i,\dot{\alpha} \alpha})$ which may be combined in an antisymmetrized product to give $\sigma^{\mu \nu} = \tfrac{1}{2}\sigma^{[\mu}\overline{\sigma}^{\nu]}$ and $\overline{\sigma}^{\mu \nu} = \tfrac{1}{2}\overline{\sigma}^{[\mu}\sigma^{\nu]}$. 

\section{Conventions for $\mathcal{N} = 2$ Hypermultiplets}
\label{appendix: hypermultiplet}

In this appendix, we study $\mathcal{N} = 2$ hypermultiplets from the perspective of bulk quantum field theories. For simplicity, we will begin by reviewing how to add a hypermultiplet describing magnetic monopoles with a $n_m$ units of magnetic charge. Because we are working with magnetically charged particles rather than electrically charged ones, the coupling of the hypermultiplet to the vector multiplet is through the S-dual variables $V_D$ and $\Phi_D$ -- these are often called magnetic variables. 

In $\mathcal{N} = 1$ language, the hypermultiplet is comprised of two chiral multiplets $(\mathcal{H}_I,\widetilde{\mathcal{H}}^I)$ with superpotential $\mathcal{W} = \sqrt{2} n_m \widetilde{H}_I \Phi_D  \mathcal{H}^I$. One infers that when the scalar field in the vector multiplet aquires a vev $\phi_D = a_D + \varphi_D$, the fields in the hypermultiplet become massive with mass $M = \sqrt{2} a_D n_m.$ A hypermultiplet with charges $(n_m,n_e)$ will generally have mass given by the BPS formula $M = \sqrt{2}|a_D n_m + a n_e|$ \cite{seiberg1994electric}. Because we are interested in chiral soft algebras, we will only examine the point in the moduli space where the hypermultiplets are massless. At this location, the cubic terms in the Lagrangian read
\begin{equation}
    \begin{split}
        \mathcal{L}_{\text{hyper, cubic}} &= -D_\mu q_I D^\mu \bar q^I + i \lambda \sigma^\mu D_\mu \bar \lambda + i \chi \sigma^\mu D_\mu \bar \chi \\
        &\hspace{20pt} + i \sqrt{2} n_m \Big(\overline{q}_{I} \psi^{I}_D \lambda - \bar{\lambda} \bar{\psi}^{I}_D q_{I}   
        - \overline{q}_I \bar{\psi}^{I}_D \bar{\chi} 
        + q_{I} \psi^{I}_D \chi + \chi \varphi_D \lambda-\bar{\lambda} {\varphi}^{\dagger}_D \bar{\chi}\Big),
    \end{split}
\end{equation}
where we have defined the covariant derivative by $D_{\mu} \equiv \partial_{\mu}- i n_m {A}_{D,\mu}$ and have expanded the $\mathcal{N} = 1$ chiral multiplets into the following component fields: an $SU(2)$ doublet of two complex scalar fields, $q_I$, satisfying the Hermiticity condition $q_I^\dagger = \overline{q}^I = \varepsilon^{IJ} \overline{q}_J$ and two singlet spinors $\lambda$ and $\chi$. 

From this Lagrangian, we can read off the following tree-level three-point amplitudes (see Appendix \ref{appendix: conventions} for conventions): 
\begin{equation}
\begin{aligned}
        \mathcal{A}_3\big( a^{A_D}_{+}(p_1) a^{q,I}(p_2) a^{\bar{q},J}(p_3) \big) &= +\sqrt{\frac{8\pi}{\text{Im} \hspace{2pt} \tau_D}}n_m \hspace{3pt}\varepsilon^{IJ} [12][13][32]^{-1}\hspace{2pt},\\
        \mathcal{A}_3\big(a^{\psi_D,J}_+(p_1) a^{q,I}(p_2) a^\lambda_{+}(p_3) \big) &= -i\sqrt{\frac{8\pi}{\text{Im} \hspace{2pt} \tau_D}}n_m \hspace{3pt}\varepsilon^{IJ}[13]\hspace{2pt}, \\
        \mathcal{A}_3\big(a^{\psi_D,J}_+(p_1) a^{\bar{q},I}(p_2) a^{\chi}_{+}(p_3) \big)  &= + i\sqrt{\frac{8\pi}{\text{Im} \hspace{2pt} \tau_D}}n_m \hspace{3pt}\varepsilon^{IJ} [13] \hspace{2pt},\\
        \mathcal{A}_3\big( a^{\overline{\varphi}_D}(p_1) a^{\lambda}_{+}(p_2) a^{\chi}_{+}(p_3) \big) &= -i\sqrt{\frac{8\pi}{\text{Im} \hspace{2pt} \tau_D}}n_m \hspace{3pt} [32]\hspace{2pt},\\
        \mathcal{A}_3\big( a^{A_D}_{+}(p_1) a^\lambda_{+}(p_2) a^\lambda_{-}(p_3) \big) &= -\sqrt{\frac{8\pi}{\text{Im} \hspace{2pt} \tau_D}}n_m \hspace{3pt} [12]^{2}[32]^{-1}\hspace{2pt},\\
        \mathcal{A}_3\big( a^{A_D}_{+}(p_1) a^\chi_{+}(p_2) a^\chi_{-}(p_3) \big) &= -\sqrt{\frac{8\pi}{\text{Im} \hspace{2pt} \tau_D}}n_m \hspace{3pt} [12]^{2}[32]^{-1}\hspace{2pt}.
\end{aligned}
\end{equation}
These are the only amplitudes which contribute to the chiral soft algebra. Note, however, that they are written in terms of dual variables rather than the usual variables from which we have written the rest of the Lagrangian. Translating to the more standard $SL_2(\mathbb{Z})$ duality frame is straightforward because both the creation operators and $\text{Im} \hspace{2pt} \tau_D$ are modular forms which respectively have weights $(\tfrac{1}{2},-\tfrac{1}{2})$ and $(-1,-1)$ (see Section \ref{sec: SL2 covariance}). It follows that:
\begin{equation}
    a_+^{A_D} = \tau^{1/2} \bar{\tau}^{-1/2} a_+^A \ ,\hspace{40pt} \text{Im} \hspace{2pt} \tau_D = \tau^{-1} \bar{\tau}^{-1} \hspace{2pt} \text{Im} \hspace{2pt} \tau = \tau^{-1} \bar{\tau}^{-1} \hspace{2pt} \frac{4\pi}{e^2} \ .
    \label{eq:aADandImTD_SL2Z}
\end{equation}
Thus, the amplitudes written in terms of more standard variables are
\begin{equation}
\begin{aligned}\label{eq:sing_3pts}
        \mathcal{A}_3\big( a^{A}_{+}(p_1) a^{q,I}(p_2) a^{\bar{q},J}(p_3) \big) &= +\sqrt{2} e \hspace{1pt}(n_e + \bar \tau n_m)\hspace{3pt}\varepsilon^{IJ} [12][13][32]^{-1}\hspace{2pt},\\
        \mathcal{A}_3\big(a^{\psi,J}_+(p_1) a^{q,I}(p_2) a^\lambda_{+}(p_3) \big) &= -i\sqrt{2} e \hspace{1pt}(n_e + \bar \tau n_m) \hspace{3pt}\varepsilon^{IJ}[13]\hspace{2pt}, \\
        \mathcal{A}_3\big(a^{\psi,J}_+(p_1) a^{\bar{q},I}(p_2) a^{\chi}_{+}(p_3) \big)  &= + i\sqrt{2} e \hspace{1pt}(n_e + \bar \tau n_m) \hspace{3pt}\varepsilon^{IJ} [13] \hspace{2pt},\\
        \mathcal{A}_3\big( a^{\overline{\varphi}}(p_1) a^{\lambda}_{+}(p_2) a^{\chi}_{+}(p_3) \big) &= -i\sqrt{2} e \hspace{1pt}(n_e + \bar \tau n_m) \hspace{3pt} [32]\hspace{2pt},\\
        \mathcal{A}_3\big( a^{A}_{+}(p_1) a^\lambda_{+}(p_2) a^\lambda_{-}(p_3) \big) &= -\sqrt{2} e \hspace{1pt}(n_e + \bar \tau n_m) \hspace{3pt} [12]^{2}[32]^{-1}\hspace{2pt},\\
        \mathcal{A}_3\big( a^{A}_{+}(p_1) a^\chi_{+}(p_2) a^\chi_{-}(p_3) \big) &= -\sqrt{2} e \hspace{1pt}(n_e + \bar \tau n_m) \hspace{3pt} [12]^{2}[32]^{-1}\hspace{2pt}.
\end{aligned}
\end{equation}
where we have used \eqref{eq:aADandImTD_SL2Z} to obtain the coupling $e\bar \tau n_m$ and further generalized the results by allowing for a non-zero electric charge in the only way consistent with duality. These three-point amplitudes are related to the coefficients $\gamma_{XYZ}$ appearing in holomorphic OPEs \cite{Himwich:2021dau}.

\bibliography{main.bib}
\bibliographystyle{jhep}

\end{document}